\documentclass[a4paper]{article}
\usepackage{fullpage}
\usepackage[round]{natbib}
\usepackage{mathptmx}
\usepackage{graphicx}
\usepackage{amsmath,stackrel,amssymb}
\usepackage[hidelinks,breaklinks]{hyperref} 
\usepackage{color} 
\usepackage{url}

\newcommand{\firstuse}[1]{\emph{#1}} 
\newcommand{\ie}{i.e.\ }
\newcommand{\eg}{e.g.\ }
\newcommand{\ignore}[1]{}
\newcommand{\unix}[1]{\texttt{#1}} 
\newcommand{\email}[1]{\href{mailto:#1}{#1}}

\providecommand{\keywords}[1]
{
  \small	
  \textbf{\textit{Keywords---}} #1
}

\begin{document}

  \title{Using Sampled Network Data With The Autologistic Actor
    Attribute Model}
  
  \renewcommand*{\thefootnote}{\fnsymbol{footnote}}
  
  \author{Alex~D.~Stivala
    \footnote{Corresponding author at: Institute of Computational Science, Universit\`{a} della Svizzera italiana, Lugano, Switzerland. Email: \email{alexander.stivala@usi.ch}}
    \and
    H.~Colin~Gallagher
    \footnote{Centre for Transformative Innovation, Faculty of Business and Law, Swinburne University of Technology, Australia} \newcounter{swin} \setcounter{swin}{\value{footnote}}
    \and
    David~A.~Rolls
\footnote{Melbourne School of Psychological Sciences, The University of Melbourne, Australia} \newcounter{melb} \setcounter{melb}{\value{footnote}}
    \and
    Peng~Wang
    \textsuperscript{\fnsymbol{swin}}
    \and
    Garry~L.~Robins  \textsuperscript{\fnsymbol{swin} \fnsymbol{melb}}
  }
  
  
  \maketitle

\begin{abstract}
Social science research increasingly benefits from statistical methods
for understanding the structured nature of social life, including for
social network data. However, the application of statistical network
models within large-scale community research is hindered by too little
understanding of the validity of their inferences under realistic data
collection conditions, including sampled or missing network data. The
autologistic actor attribute model (ALAAM) is a statistical model
based on the well-established exponential random graph model (ERGM)
for social networks. ALAAMs can be regarded as a social influence
model, predicting an individual-level outcome based on the actor's
network ties, concurrent outcomes of his/her network partners, and
attributes of the actor and his/her network partners. In particular,
an ALAAM can be used to measure contagion effects, that is, the
propensity of two actors connected by a social network tie to both
have the same value of an attribute. We investigate the effect of
using simple random samples and snowball samples of network data on
ALAAM parameter inference, and find that parameter inference can still
work well even with a nontrivial fraction of missing nodes. However it
is safer to take a snowball sample of the network and estimate
conditional on the snowball sampling structure.
\end{abstract}


\section{Introduction}
\label{sec:intro}

An autologistic actor attribute model (ALAAM)
\citep{robins01b,daraganova13} is a statistical model based on the
widely used exponential random graph model (ERGM) for social networks
\citep{frank86,robins01,robins07,lusher13}. An ERGM can be used to
model social networks, predicting the presence of a tie between two
actors based on other ties (structural properties of the network) and
attributes of the actors (nodes) themselves. By contrast, an ALAAM can
be used to predict an attribute of an actor, based on the actor's ties
to other nodes in the network, as well as attributes of the
actor and its network partners. In this way it is similar to logistic
regression, but, unlike logistic regression or similar statistical
techniques, specifically does \emph{not} assume independence of the
predicted attributes across actors --- an actor's outcome attribute may depend also
on those of its neighbors in the network. Hence an ALAAM may be used
as a model of social influence, examining how some attribute of an actor
in a network is affected by his or her position in the network and the
attributes of other actors in the network.

Although ALAAMs are not widely cited (they are not included in
\citet{silk17}, for example) they have been used to model the
acquisition of descriptive norms as social category learning in social
networks \citep{kashima13}, spatial and network influence processes on
unemployment \citep{daraganova13b}, and the performance of researchers
\citep{letina16,letina16b}.

As a result, ALAAMs have not been as widely studied as the
better-known multiple regression quadratic assignment procedure
(MRQAP) \citep{krackhardt88b,dekker07} or network autocorrelation model
\citep{ord75,cliff81,doreian81,anselin90,friedkin90,leenders02}. It is
known that the network autocorrelation model has a systematic negative
bias in the estimation of the network effect under almost all
conditions \citep{mizruchi08,neuman10}, and \citet{wang14} present a
detailed analysis of the statistical power and type I error rate of
the network autocorrelation model.  More recently, \citet{sewell17}
proposes a method to use network autocorrelation with egocentric
network samples, where only ties directly involving randomly sampled
actors are considered, and \citet{dittrich17,dittrich19} propose a
Bayesian approach to the network autocorrelation model.

ALAAMs, being the social influence counterpart of the social selection
ERGM model, share the latter's flexibility in that a wide variety of
effects can be included in the model (subject to dependence
assumptions). This makes them potentially much larger in scope (but
more complex) than models such as network autocorrelation, which only
estimate a single network effect. In this paper we investigate the
effect of missing data and network sampled data on the inference of
ALAAM parameters. We examine two forms of sampling: first, data
missing at random, as would occur when taking a simple random sample
from the network, and, second, network samples obtained via snowball
sampling.

A practical motivation for this study is the manner in which social
influence models may be applied to epidemiological studies of health
outcomes in large-scale community samples (\eg
\citealt{gibbs13}). Conventionally, these studies often use
cross-sectional data to examine the prevalence of health conditions
(\eg \citealt{dirkzwager06}) or probable mental health conditions
across a given population (\eg \citealt{kessler02,bryant14}).  In such
settings, a methodological priority is placed on obtaining some form
of random sample: given the impracticality of interviewing an entire
general population, random sampling allows one to make valid
inferences based on a representative sample of individuals drawn from
that population. However, the individualistic assumptions upon which
such research designs are usually based are increasingly at odds with
conceptual frameworks of human behavior which appeal to the
structured nature of social life. For instance, health researchers and
other social scientists increasingly seek to understand health and
other individual outcomes as embedded within larger social systems of
interconnected actors, employing an array of socio-structural
concepts, including social capital \citep{desilva05,kawachi13},
community resilience \citep{norris08}, and community wellbeing
\citep{gibbs15}.

Social network research methods offer a resolution to this tension,
particularly statistical models for social networks, illuminating the
multifaceted role that network ties play in the prevalence of health
issues, including processes of social support, social selection, and
social influence and diffusion \citep{valente10}. However, in many
instances, a complete census of a network may be prohibitively
expensive and difficult: it may be the case that not every member of
the population can be identified, much less recruited into a
study. Complicating matters is that network researchers cannot rely on
random sampling techniques, which have been designed to mitigate the
very thing which is of primary interest: social interdependence among
actors \citep{robins15}. We are thus left with an important dilemma:
sociocentric research methods are often difficult to execute properly,
depending on the research setting, yet the questions that these
methods can address become no less important, such as the prevalence
and spread of mental health issues across a disaster-affected area. A
central issue is therefore the degree to which network data may be
missing, either randomly, or by some design (\ie sampled), and still
yield valid inferences.

As a result, a considerable amount of research has been directed at
identifying the impact of different forms of missing data, or the
sampling of network data, and adequately accounting for it within
statistical models. Work examining the effect of missing data on
social network parameter estimation goes back at least to
\citet{holland73}.  More recently, \citet{kossinets06}
examines the effect of missing data on estimating network statistics
of a bipartite graph.
\citet{smith13} investigates the effects of nodes missing at random,
and \citet{smith17} the effects of non-random missing data, on network
centrality, degree homophily, and topology measures.  \citet{robins04}
investigate the effect of missing data on ERGM parameter
estimation. \citet{koskinen13a} propose a Bayesian technique to handle
missing data in ERGM parameter estimation. Previous work that has
examined the effect of snowball sampling on the estimation of social
network properties includes
\citet{handcock10,thompson00,illenberger12,pattison13}.
\citet{handcock10}, \citet{pattison13}, and \citet{stivala16} describe
methods to estimate ERGM parameters from snowball samples, while
\citet{thompson00} describe a maximum likelihood estimator of
population network parameters given a snowball sample, and
\citet{illenberger12} describe estimators of topological network
parameters using snowball samples. Nonetheless, efforts to describe
the appropriate sampling techniques for network research are still at
an early stage, and the application of sociocentric (whole-network)
methods to questions of health remain limited to relatively small,
well-delineated systems in which the collection of near-complete
whole-network data is feasible, such as networks of students within a
high school (\eg \citet{kiuru12}).

However, research has still only partially addressed this question of
missing and sampled network data. Previous work on missing network
data and the sampling of network data pertains mostly to
network statistics (\eg \citealt{borgatti06b,smith13,smith17}), or
social
selection models, namely ERGM, in which one aims to predict tie
formation based on an exogenous set of actor attributes, and
endogenous processes of network tie formation. By contrast, no
existing work examining the effect of missing or sampled data on its
social influence counterpart, ALAAM, in which the aim is to predict an
individual outcome based on an exogenous set of network ties, an
exogenous set of personal attributes, and endogenous processes of
social influence or diffusion. Thus, social influence models are
advantageous in addressing the same questions as the conventional
regression models used within epidemiological studies (\eg logistic
regression), while at the same time taking account of network
interdependencies among these outcomes in a principled manner. This
article therefore aims to commence research on whether ALAAM
represents a valid socio-structural approach with missing and sampled
network data, which represent a more realistic set of circumstances in
many empirical settings.

\section{Autologistic Actor Attribute Models (ALAAMs)}
\label{sec:alaam}

ALAAMs were first described in \citet{robins01}, modeling the probability of attribute $Y$ (a vector of binary attributes) given the network $X$ (a matrix of 0-1 tie variables). The model can be expressed in the
form \citep{daraganova13}:
\begin{equation}
  \Pr(Y = y \vert X =x) = \frac{1}{\kappa(\theta_I)}\exp\left(\sum_I \theta_I z_I(y,x,w)\right)
\end{equation}
where $\theta_I$ is the parameter corresponding to the network-attribute
statistic $z_I$, in which the ``configuration'' $I$ is defined by a combination
of dependent attribute variables $y$, network variables $x$, and
actor covariates $w$, and
$\kappa(\theta_I)$ is a normalizing quantity which ensures a proper
probability distribution.

The ALAAM predicts the outcome variable $Y$, taking into account
network dependencies in a principled manner which is not possible with
standard logistic regression.  Assumptions about which attributes $Y$
are independent, and therefore which configurations $I$ are allowed in
the model, determine the class of the model.  In the simplest case, in
which any two attribute variables $Y_i$ and $Y_j$ are assumed to be
independent, the only possible configuration is a single node.  Then
there are no network effects, and the model reverts to standard
logistic regression \citep[p. 105]{daraganova13}.

The simplest network dependence assumption is that an attribute
variable $Y_i$ is conditionally dependent on  network tie $X_{jk}$
if and only if $\{i\} \cap \{k,j\} \neq \emptyset$, that
is, if and only if the actor $i$ is one end of the tie $X_{jk}$.
Hence the configurations $I$ allowed in this class of model include stars
\citep[p. 107]{daraganova13}, as well as the \firstuse{contagion}
effect, that is, the propensity of two nodes with a tie between them
to both have the attribute $Y$.

We will use this dependence assumption, and also include covariate effects,
in which other attributes of an actor influence its outcome attribute $Y$,
just as they would in standard logistic regression if we did not have the
network effects. 

\begin{figure}
  \centering
  \includegraphics[scale=0.2]{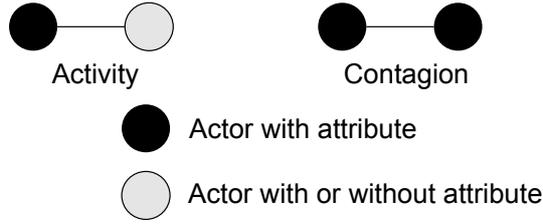}
  \caption{The structural ALAAM configurations used.}
  \label{fig:attr_configurations_alaam}
\end{figure}

The ALAAM configurations we will use are:

\begin{description}
  \item[Attribute density] The number of nodes with attribute $Y$;
  \item[Activity] Presence of a tie at a node with attribute $Y$. That is, whether having attribute $Y$ is associated with having a tie to others;
  \item[Contagion] The propensity for two nodes with a tie between them to both have attribute $Y$;
  \item[Binary] The propensity for a node to have attribute $Y$ based on another binary attribute $U$;
  \item[Continuous] The propensity for a node to have attribute $Y$ based on its continuous attribute $V$.
\end{description}
The structural configurations are shown in Figure~\ref{fig:attr_configurations_alaam}.

By estimating the parameters $\theta_I$, we can make inferences as to
whether or not the corresponding effects are significant, given the
other effects included in the model. For example, if the Contagion
parameter is positive and significant, it means that, in the network
in question, the number of directly connected actors that both have
the outcome attribute is more than would be expected to occur
by chance, given the other effects in the model. The Binary and Continuous parameters therefore have exactly
the same meaning as they would in standard logistic regression, with
the exception that, in our model which includes network effects, their
value is estimated while controlling for the 
structural effects.

In this paper we examine the effect of missing data
and snowball samples on the validity of these inferences.

\section{Snowball Sampling}
\label{sec:snowball}

Snowball sampling \citep{coleman58,goodman61} is a method to generate
a sample of nodes in a network by ``link tracing'', that is, when a node
is in the sample, obtaining further nodes by ``following'' links (ties)
from the node in question. There are many variations and other techniques
related to snowball sampling, and a useful set of papers
\citep{goodman11,heckathorn11,handcock11} clarifies the distinctions 
between them.

\begin{figure}
    \centering
    \includegraphics[width=\textwidth]{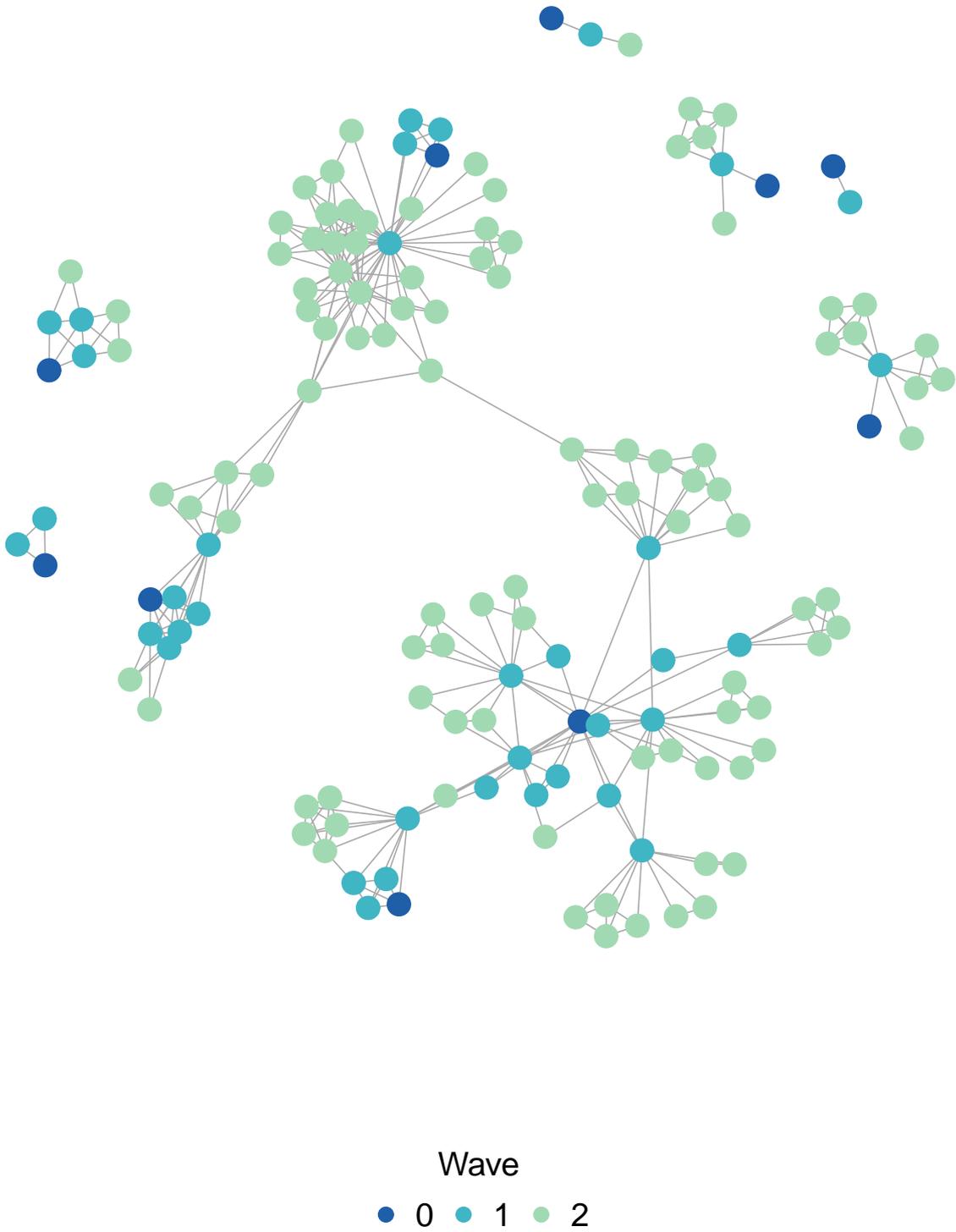}
    \caption{A two wave snowball sample, obtaining 146 nodes from a 1589 node network.}
    \label{fig:subgraph24}
\end{figure}

To obtain a snowball sample from a network, the first step is to take
a random sample of some number of nodes, called the \firstuse{seed nodes} or
\firstuse{wave 0} of the sample. Then wave 1 of the sample consists of
all nodes that have a tie to a node in wave 0, but are not themselves
in wave 0. Wave $l$ of the snowball sample consists of all nodes that
have a tie to a node in wave $l-1$ of the sample, but are not
themselves in waves $0, \ldots, l-1$. The $l$-wave snowball sample is
then the set of all nodes thus obtained, and the ties between them ---
that is, the subgraph induced by the nodes in the sample.
This is illustrated in Figure~\ref{fig:subgraph24}.

We use two forms of snowball sampling in this paper. In the first
form, at most a fixed number $m$ of ties is followed from a node to
obtain more nodes in the sample. This is the ``traditional'' form
described in \citet{goodman61}, and is the situation in a
survey design where respondents are asked to name up to $m$ friends, for
example. This is also known as a \firstuse{fixed choice} design, and
has also been described as \firstuse{degree censoring}
\citep{kossinets06} at $m$.

In the second form, there is no such limit $m$ and all ties
from a sampled node are followed. This is equivalent to $l$ steps of
breadth-first search (BFS) in the network, and is therefore also known
as \emph{BFS sampling}, and is used for example in
\citet{newman03,lee06,kurant11,pattison13,stivala16}. In the graphs shown in
the results in the following sections, we denote this form with
``$m=\mbox{Inf}$'' since it is equivalent to sampling with an infinite
value of the fixed maximum number of ties $m$.

The snowball sampling parameters that govern the structure and size 
of a snowball sample are then: the number of waves, the number of seed nodes,
and $m$, the maximum number of ties to follow (when $m=\mbox{Inf}$ 
it is BFS sampling, otherwise it is a fixed choice snowball sampling design).

\begin{figure}
    \centering
    \includegraphics[width=\textwidth]{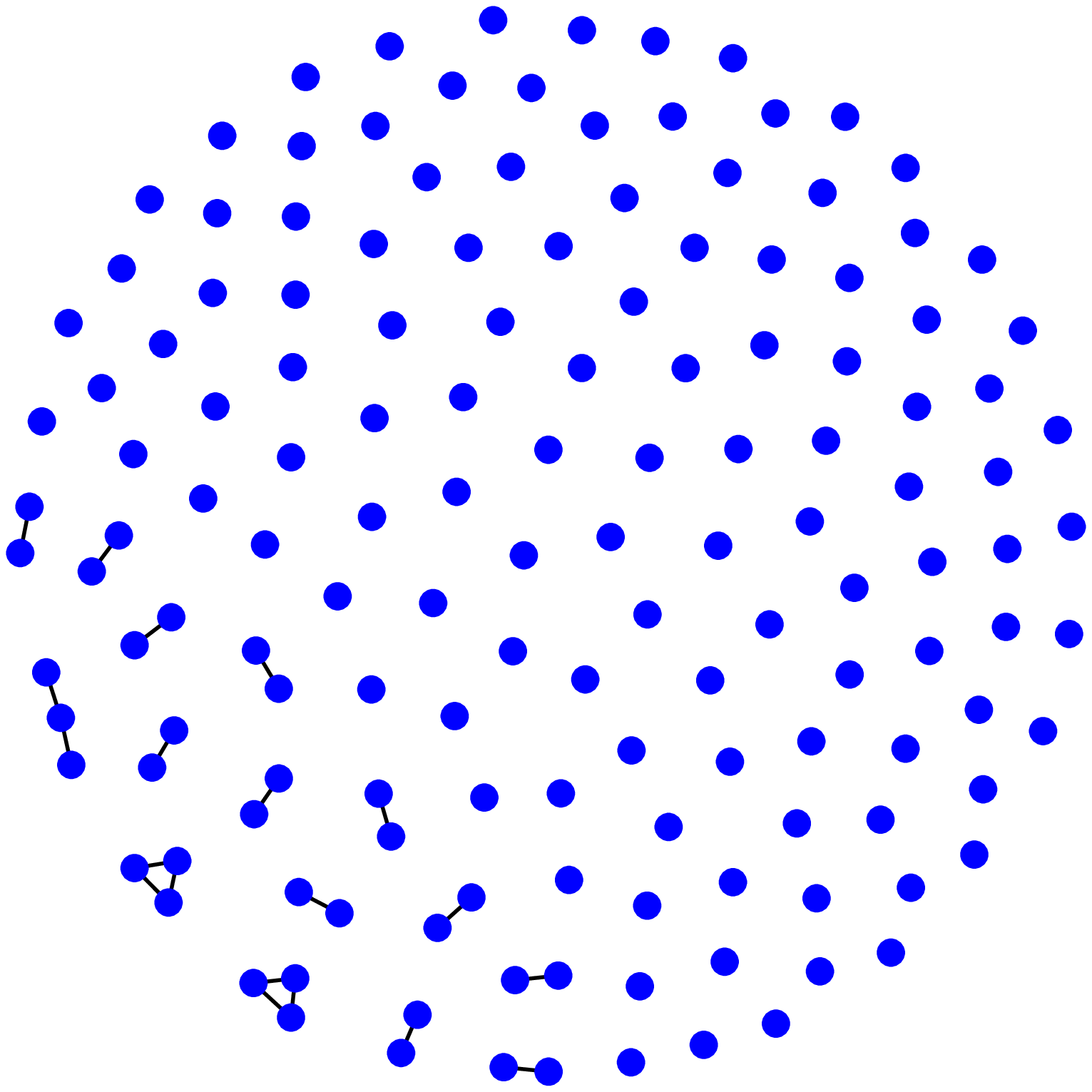}
    \caption{A network sample obtained as the subgraph induced by a random sample of 146 nodes from a 1589 node network.}
    \label{fig:randomgraphsample}
\end{figure}

\begin{figure}
    \centering
    \includegraphics[width=\textwidth]{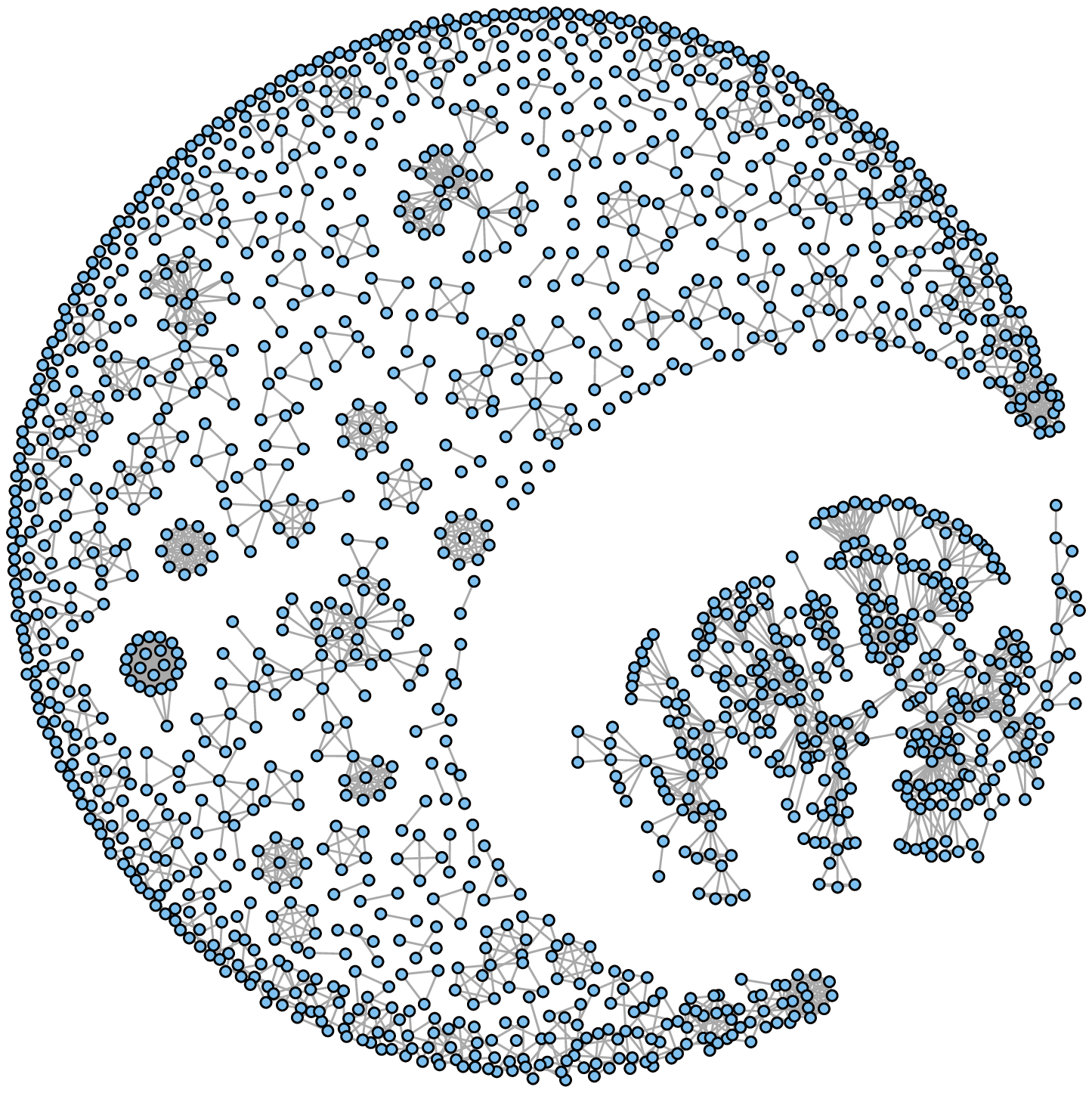}
    \caption{The network science collaboration network, $N=1589$.}
    \label{fig:netscience}
\end{figure}

The advantage of snowball sampling in obtaining a sample retaining
network structural information is illustrated by
Figures~\ref{fig:subgraph24} and
\ref{fig:randomgraphsample}. Figure~\ref{fig:subgraph24} is a two-wave (and
ten seed nodes) snowball sample from the network science collaboration
network (Figure~\ref{fig:netscience}) \citep{newman06}. This sample includes 146 nodes (of
the original 1589), \ie approximately 9\% of the nodes in the original
network. Figure~\ref{fig:randomgraphsample} is the network resulting
from sampling exactly the same number of nodes at random from the
original network. It is clear that this retains very little of the
original network structure (most nodes are isolates, and the largest
structures are cliques of size three, \ie triangles) compared to the
snowball sample.

\subsection{Conditional Estimation}

ALAAM parameters are usually (when a full network is available)
estimated by maximum likelihood estimations (MLE) using a stochastic
approximation algorithm \citep{snijders02,lusher13}. In short, this
involves a Markov chain Monte Carlo (MCMC) procedure, whereby, at each
iteration, the outcome variable for a randomly chosen actor is
toggled, and the resulting changes in the sufficient statistics of the
model are used to compute the acceptance probability in a
Metropolis--Hastings algorithm to simulate a sequence of simulated
outcome vectors. These are then used in an iterative algorithm to
estimate ALAAM parameters by stochastic approximation.

When parameters are to be estimated from a snowball sample, however, a
conditional estimation algorithm, as proposed in the similar situation
for ERGM estimation \citep{pattison13} is used. Conditional estimation
for snowball sampled data in the context of ALAAM parameter estimation
is described in \citet{daraganova09,daraganova13b,kashima13}. The
essential idea is that the estimates are conditional on the fixed
network and on the outcome attributes fixed in the outermost wave
(wave 2 in the two-wave snowball sample illustrated in
Figure~\ref{fig:subgraph24}), as in this last wave of the snowball
sample, there is no data about their network nominations (other than
those already in earlier waves), and so the outcome variable of these
nodes is made exogenous to the model. In concrete terms, this means
that in the MCMC procedure, only nodes in the innermost waves have
their outcome variable toggled.

\section{Implementation}
\label{sec:impl}

ALAAM estimation is done using IPNet \citep{wang09}, with the
estimations, one for each of the $N_A = 100$ simulated ALAAMs, run in
parallel on a Linux compute cluster using \unix{GNU parallel}
\citep{tange11}. Only converged parameter estimates, defined as those
with a reported $t$-ratio with a magnitude less than 0.1, are
included.  A parameter estimate is considered significant if its
magnitude is more than twice its standard error.

Scripts for sampling in networks, visualization, and bootstrap
error estimation are written in R \citep{R-manual} using the
\unix{igraph} package \citep{csardi06}. Graphs are generated with the R
\unix{ggplot2} package \citep{wickham09}; locally weighted polynomial
regression curves are generated using the \unix{loess} function with
default parameters.


In graphs and inference, 95\% confidence interval are used.
Confidence intervals on RMSE plots are computed by the non-parametric
bias-corrected and accelerated (BCa) method \citep{efron87,davison97}.
This method adjusts for bias and skewness in the bootstrap
distribution.  The bootstrap replicates are constructed by taking
random resamples of size $N_A$ with replacement from the $N_A$
estimates (one for each simulated ALAAM).  Bootstrap confidence
intervals are estimated with 20000 replicates using the R \unix{boot}
package \citep{davison97}.

The confidence interval shown for estimated type I and type II
error rates is the 95\% confidence interval for the binomial proportion
computed using the Wilson score interval \citep{wilson27}.

\section{Simulation Studies}
\label{sec:simulation}

To evaluate the behavior of ALAAM estimation with missing data, we require
ALAAMs with known parameters. This is done by simulating an ALAAM (that is, 
generating the outcome variable from specified ALAAM parameters) on a fixed
network. Hence we first need a network to use. We use an undirected network with 500 nodes. This network was taken
as a single sample from a sequence of simulated networks generated by 
PNet \citep{wang09} with the ERGM edge, alternating $k$-star, 
alternating $k$-triangle, 
and alternating 2-path parameters equal to -4.0, 0.2, 1.0, and -0.2,
respectively. These parameters are chosen to generate ``reasonable''
network statistics, in line with parameters estimated from empirical networks,
and are the same as those used in \citet{pattison13}.
The networks are sampled from a Markov chain Monte Carlo (MCMC) simulation,
with sufficient burn-in (of the order of $10^7$ iterations) to ensure initialization effects are minimized.

As well as the simulated network, we use two empirical networks, as used in a study of respondent-driven sampling \citep{goel10}. The
first is the ``Project 90'' network, a sexual contact network of
high-risk heterosexuals in Colorado Springs
\citep{potterat04,woodhouse94,klovdahl94,rothenberg95}.
The second is a friendship network from the National Longitudinal Study
of Adolescent Health (``Add Health'') \citep{addhealth,moody01}.
As in \citet{goel10}, we use only the giant components of these networks.
Descriptive statistics of the networks are shown in Table~\ref{tab:graphstats}.


Each node in each network has a binary and a continuous attribute. 
The binary attribute is assigned the positive value for 50\%  of the nodes,
chosen at random. For the continuous attribute, the attribute value $v_i$ at each node $i$ is 
$v_i \stackrel{iid}{\thicksim} N(0,1)$. 

The simulated ALAAMs are generated with the
parameters shown in Table~\ref{tab:simdata}.  These parameters were
chosen so that approximately 15\% of nodes have a positive outcome
variable.  For each set of parameters in this table, a set of 100
outcomes is sampled from the distribution with those parameters using
IPNet \citep{wang09}. The outcomes are sampled from a MCMC
distribution with sufficient burn-in ($10^6$ iterations for the 500 node simulated network, $10^7$ iterations for the Add Health network, and $10^8$ iterations for the Project 90 network) to ensure
initialization effects are minimized, and the samples are taken far
enough apart ($10^5$ iterations for the simulated network and $10^7$ iterations for the larger empirical networks) to ensure that they are essentially
independent. The proportion of nodes with a positive outcome variable
is shown in Table~\ref{tab:graphstats} --- it is close to 15\% on average.
This proportion was selected to correspond roughly to rates of mental
health conditions within a disaster-affected population (\eg
\citealt{bryant14}).

\begin{table}
  \begin{center}
\begin{tabular}{lrrrrrrrr}
\hline
Network & N  &   Components &  Mean  & Max.      &    Density & Clustering & \multicolumn{2}{c}{Positive outcome \%}\\
        &     &             &   degree       &     degree     &            &  coefficient          & mean & s.d.             \\
\hline
ERGM & 500 & 5 & 4.90 & 11 & 0.00983 & 0.10347 & 15 & 2.19\\
Project 90 & 4430 & 1 & 8.31 & 159 & 0.00188 & 0.34332 & 15 & 0.34\\
Add Health & 2539 & 1 & 8.24 & 27 & 0.00324 & 0.14189 & 15 & 0.54\\
\hline
\end{tabular}
  \end{center}
\caption{Network statistics of the simulated (ERGM) and empirical networks.}
\label{tab:graphstats}
\end{table}

\begin{table}
  \begin{center}
  \begin{tabular}{lrrrrrr}
    \hline
     Network & N & Density & Activity & Contagion & Binary & Continuous \\
    \hline
    ERGM & 500 &  -7.20& 0.55& 1.00& 1.20 & 1.15    \\
    Project 90 & 4430 & -15.0 & 0.55 & 1.00 & 1.20 & 1.15 \\
    Add Health & 2539 & -12.5 & 0.55 & 1.00 & 1.20 & 1.15 \\
    \hline
  \end{tabular}
  \end{center}
\caption{Parameters of the simulated ALAAMs.}
\label{tab:simdata}
\end{table}

\begin{figure}
  \centering
  \includegraphics[width=\textwidth]{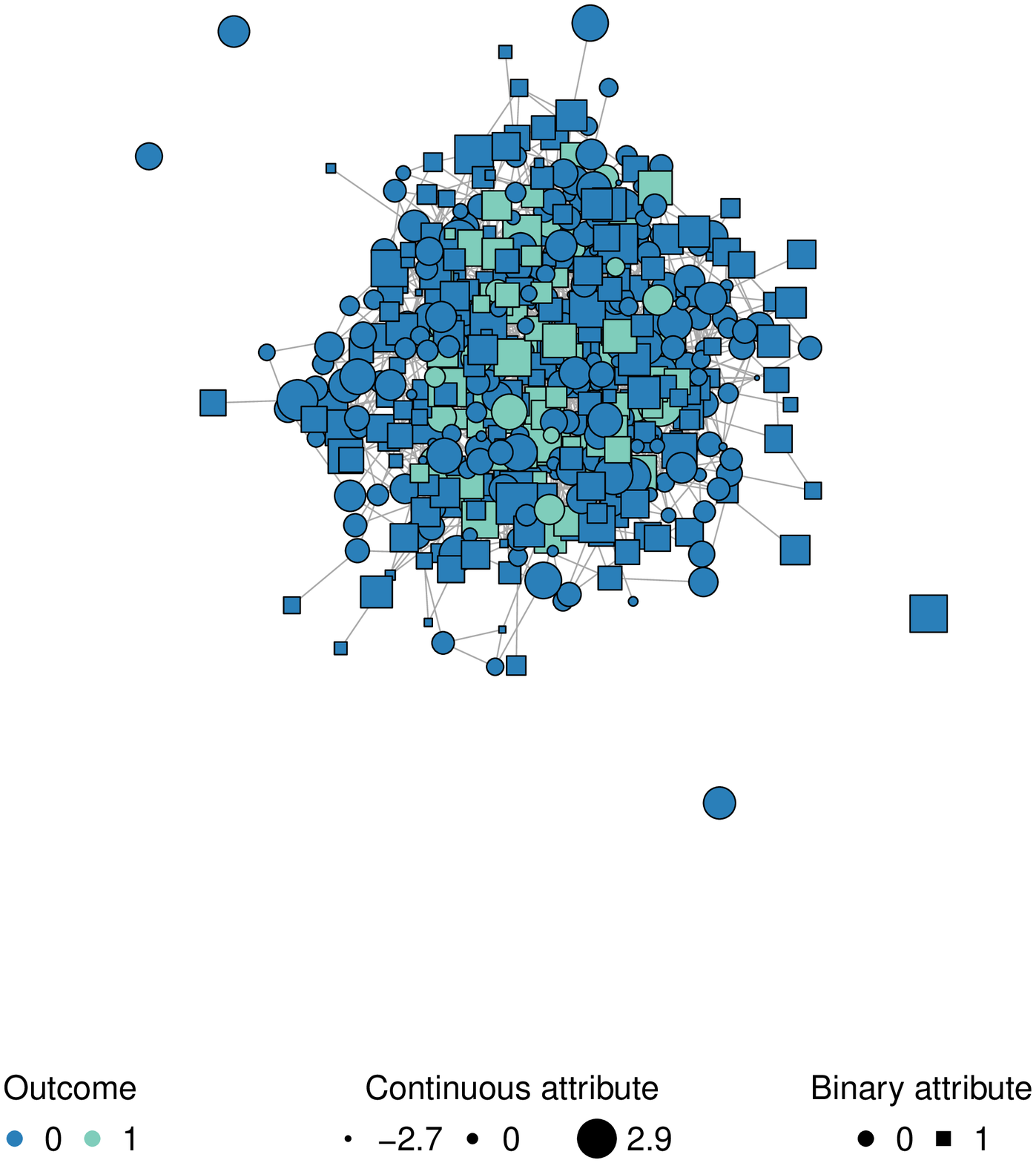}
  \caption{The simulated network $N=500$ with nodes shaped according
    to the binary attribute and node size proportional to the
    continuous attribute. Nodes with
    the positive outcome variable are colored light blue/green (16.2\%
    of the nodes) and the others dark blue.}
  \label{fig:samplegraph}
\end{figure}

Figure~\ref{fig:samplegraph} shows the simulated 500 node network with nodes
colored according to the outcome variable for one sample from the ALAAM
simulations.

\subsection{Simulation Study 1: Random Sampling of Nodes}

In the first study we investigate the effect of sampling nodes
independently at random from the network, which we refer to as
random node sampling, or
``simple random sampling''.  The network then used for estimation is
the subgraph of the original graph induced by the selected (not
omitted) set of nodes. In other words, when a node is not included,
all edges connected to that node are also not in the resulting
network.  We estimate the ALAAM parameters from this sampled
subnetwork, comparing the results to the estimation of the full
network.

\begin{figure}
  \includegraphics[width=\textwidth]{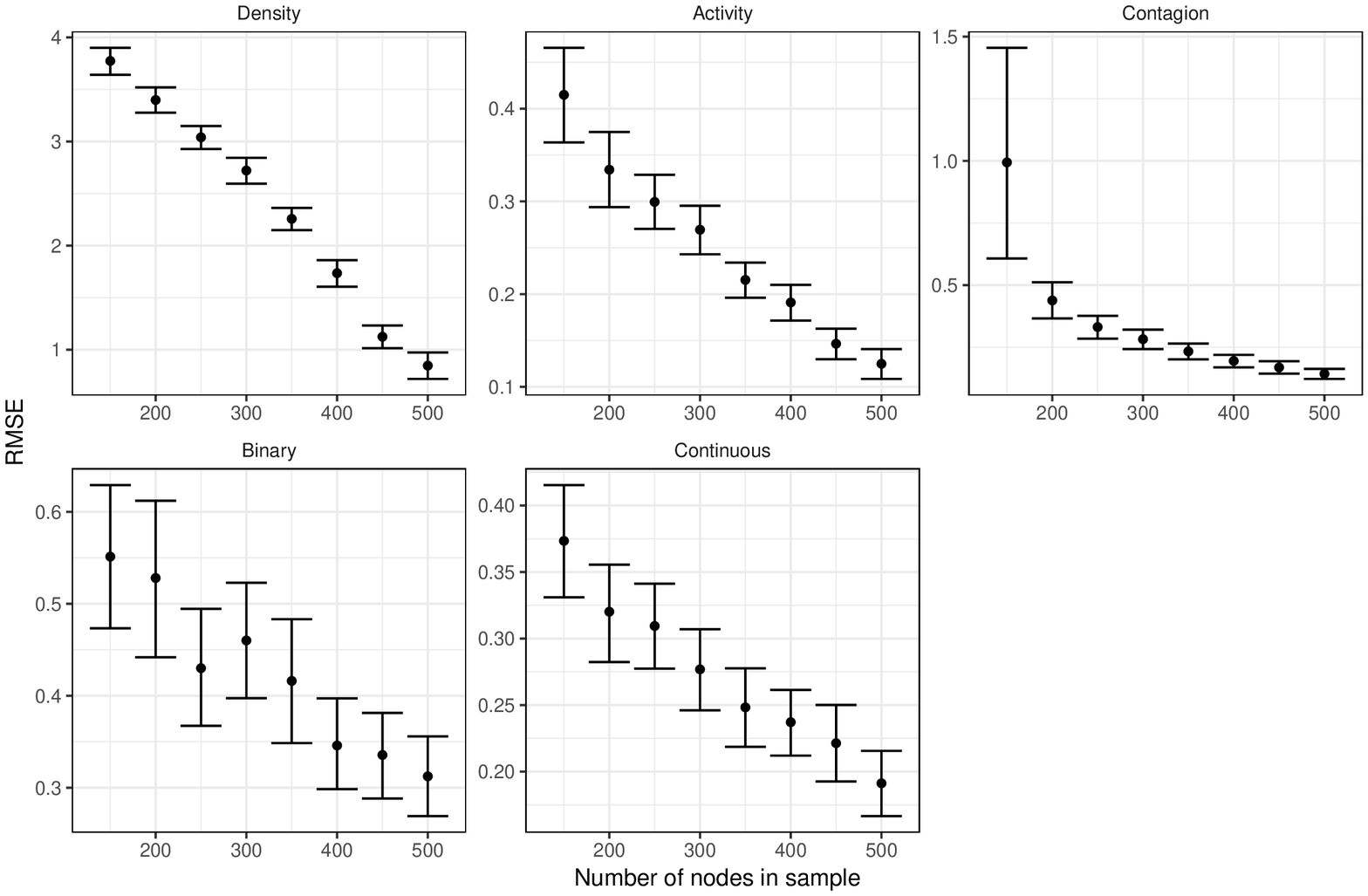}
\caption{Effect on root mean square error of random node sampling. Simulated ERGM network $N=500$.}
 \label{fig:rmse_random}
\end{figure}

Figure~\ref{fig:rmse_random} shows the effect of random node sampling
on the root mean square error (RMSE) in ALAAM parameter estimation.
The RMSE is the square root of the mean squared difference between the
estimate and the true value. (Note that in this figure the RMSE values
are on different scales for each parameter). The RMSE decreases steadily
as more the sample size increases, although for the Contagion parameter there
appears to be a larger jump from 100 to 200 nodes.

\begin{figure}
  \includegraphics[width=\textwidth]{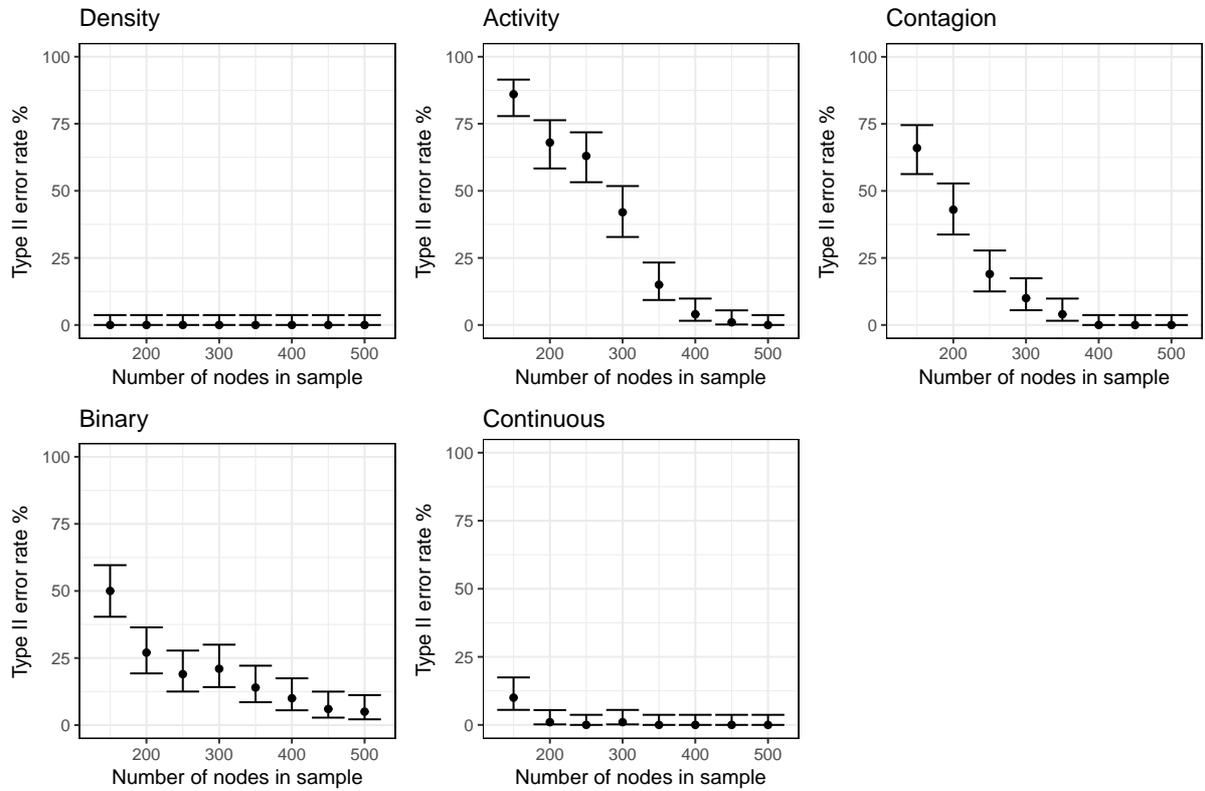}
  \caption{Effect on type II error of network random node sampling. Simulated ERGM network $N=500$.}
  \label{fig:random_fnr}
\end{figure}

Figure~\ref{fig:random_fnr} shows the effect of random node sampling
on the type II error rate, that is, the false negative rate, in ALAAM
parameter inference. This is the percentage of experiments (over the
100 simulated ALAAMs) in which the estimate has the wrong sign or the
confidence interval covers zero (so we cannot reject the null
hypothesis that the parameter for the effect is zero). With the
exception of the Activity parameter, the type II error rate does not
increase greatly even when only 300 nodes (60\% of the total) are
sampled. Hence even if we can only sample (at random) 60\% of the
network, ALAAM inference still has good power on effects other than
Activity. Even this low power relative to the other effects could be
because the Activity parameter in our simulations has a small
magnitude compared to the other effects.

\begin{figure}
  \includegraphics[width=\textwidth]{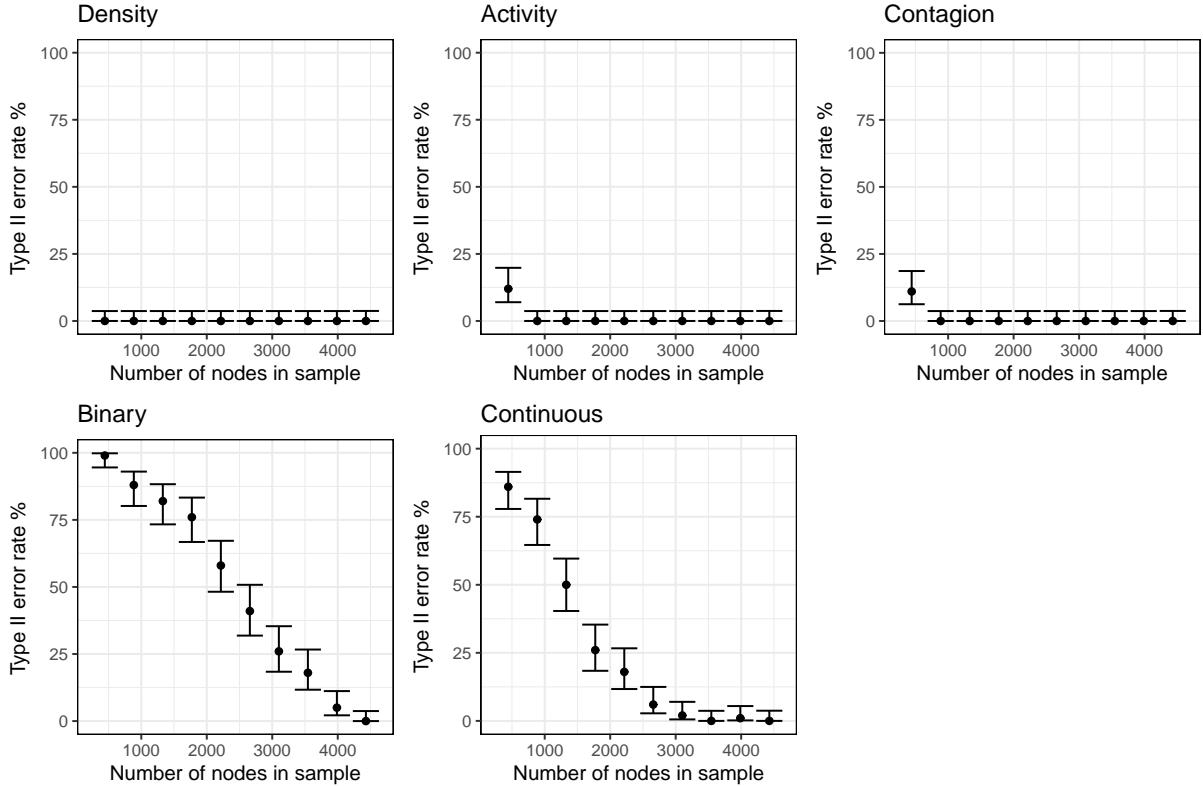}
  \caption{Effect on type II error rate of random node sampling. Project 90 network $N=4430$.}
  \label{fig:project90_random_fnr}
\end{figure}

\begin{figure}
  \includegraphics[width=\textwidth]{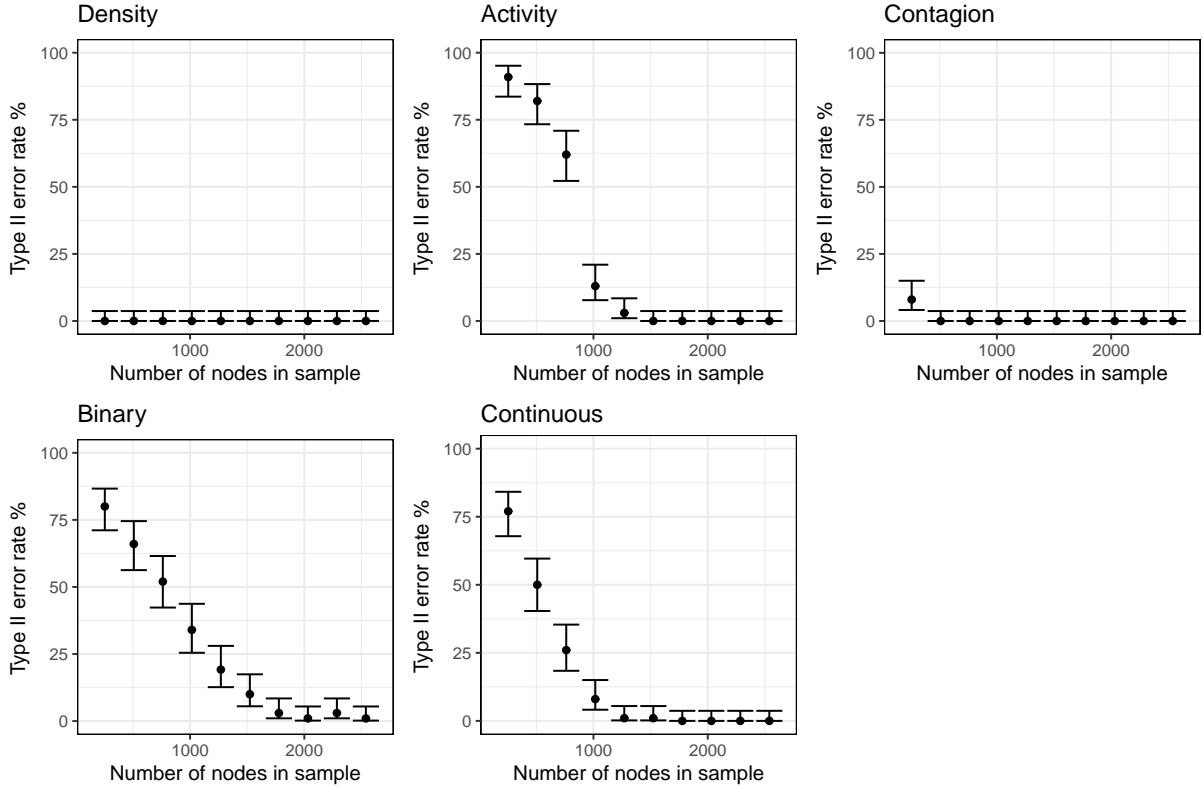}
  \caption{Effect on type II error rate of random node sampling. Add Health network $N=2539$.}
  \label{fig:addhealth_random_fnr}
\end{figure}

Figure~\ref{fig:project90_random_fnr} for the Project 90 network and
Figure~\ref{fig:addhealth_random_fnr} for the Add Health network, show
the effect of increasingly larger random node sampling sizes on the
type II error rate. In both networks, Density and Contagion have very
low type II error rates over the whole range of sample sizes, while
the rates for Binary and Continuous decrease slowly. The behavior of
the Activity parameter, however, is quite different. For the Project
90 network (Figure~\ref{fig:project90_random_fnr}) the type II error
rate is very low for the entire range of sample sizes, while for the
Add Health network (Figure~\ref{fig:addhealth_random_fnr}) the error
rate is very high for less than 1000 nodes, and rather abruptly
becomes very low for larger sample sizes. This in turn is different
from the behavior on the simulated network where the type II error
rate on the Activity parameter declines more smoothly with increased
sample size (Figure~\ref{fig:random_fnr}).

In order to measure the type I error rate in inference (false positive
rate), for an effect, we require simulated ALAAMs which do not have
that effect present (its parameter is zero).  So for each of our ALAAM
effects (except Density, which if zero results in almost all nodes
having a positive outcome variable), we simulate another set of 100
ALAAM outcomes in which the corresponding parameter is set to
zero. This allows us to test for false positives with respect to this
zero effect, that is, the percentage of experiments in which, for an
estimate of a zero effect parameter, the confidence interval does not
include zero.  The other parameters are retained at their values shown
in Table~\ref{tab:simdata}, except in the case of the Activity effect,
which when it is zero, results in less than 1\% of the nodes having a
positive outcome variable, and so the Density parameter was increased
to -4.0 for the simulated ERGM network, -7.0 for the Project 90
network, and -6.0 for the Add Heath network, to obtain a reasonable
number of nodes with a positive outcome variable.

\begin{figure}
  \includegraphics[width=\textwidth]{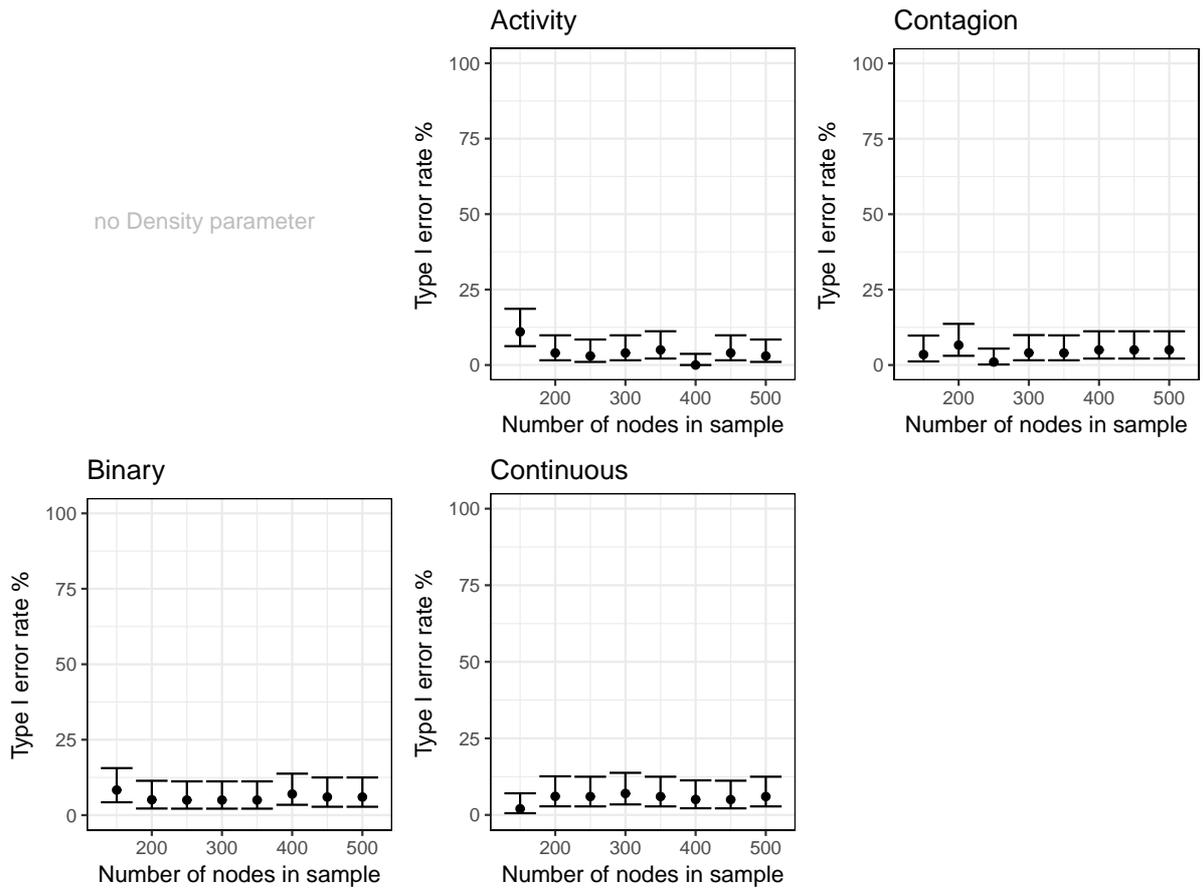}
  \caption{Effect on type I error rate of random node sampling. Simulated ERGM network $N=500$.}
  \label{fig:random_fpr}
\end{figure}

The type I error rates obtained in this way are shown in
Figure~\ref{fig:random_fpr} for the simulated network, which shows
that even when the sample size is only 100 nodes (20\% of the nodes),
the type I error rate does not increase significantly on any of the
ALAAM effects tested.  It is notable that the type I error rates for
contagion and the other attribute predictors hardly change as the
sample becomes a smaller proportion of the network, illustrating that,
at least for this example, the inference of a ``significant'' effect
is robust for a sample that is only 20\% of the entire network.

Only 1.7\% of ALAAM estimations just described do not converge,
consisting almost entirely of those with very small sample sizes.

\begin{figure}
  \includegraphics[width=\textwidth]{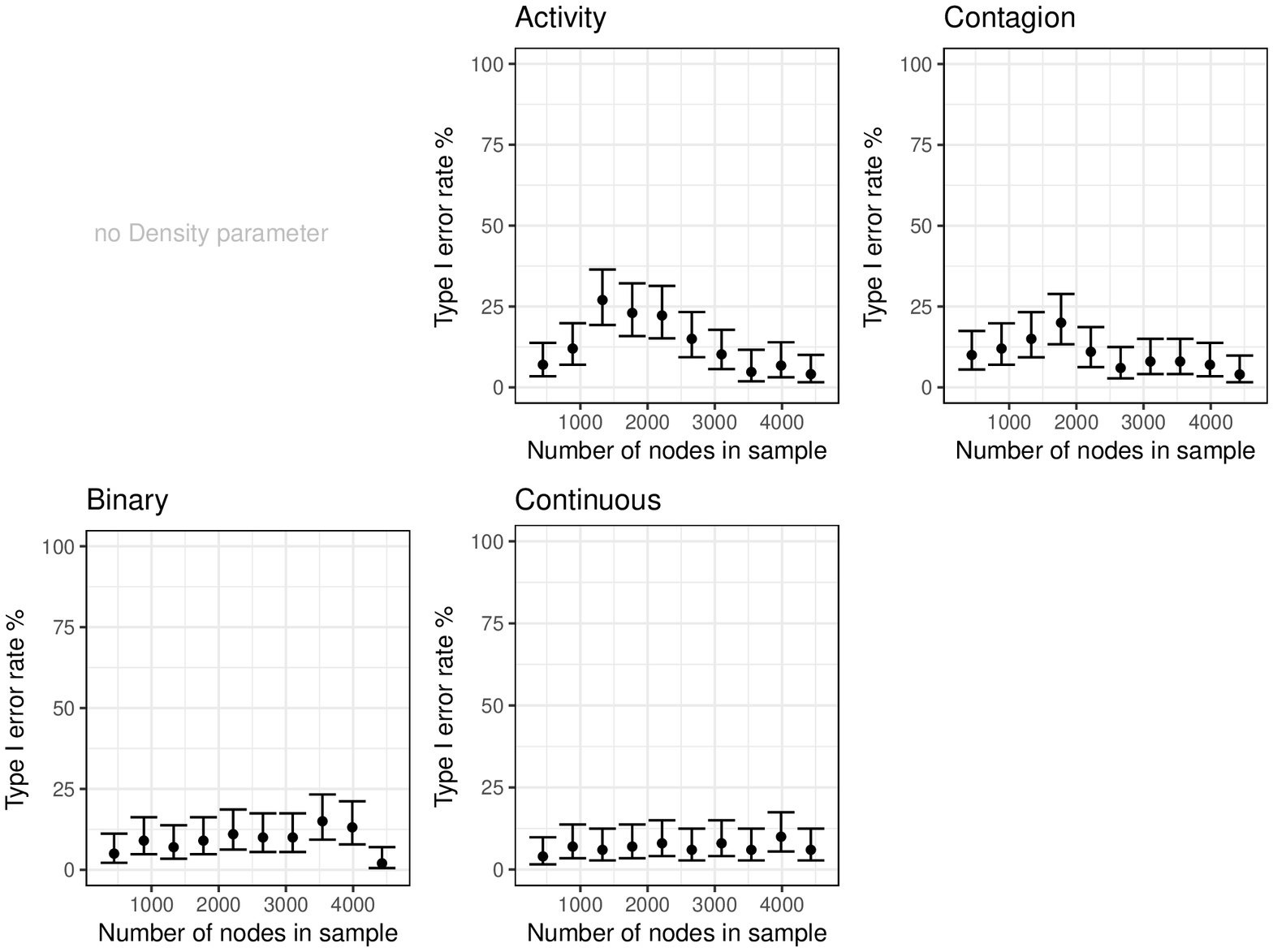}
  \caption{Effect on type I error rate of random node sampling. Project 90 network $N=4430$.}
  \label{fig:project90_random_fpr}
\end{figure}

\begin{figure}
  \includegraphics[width=\textwidth]{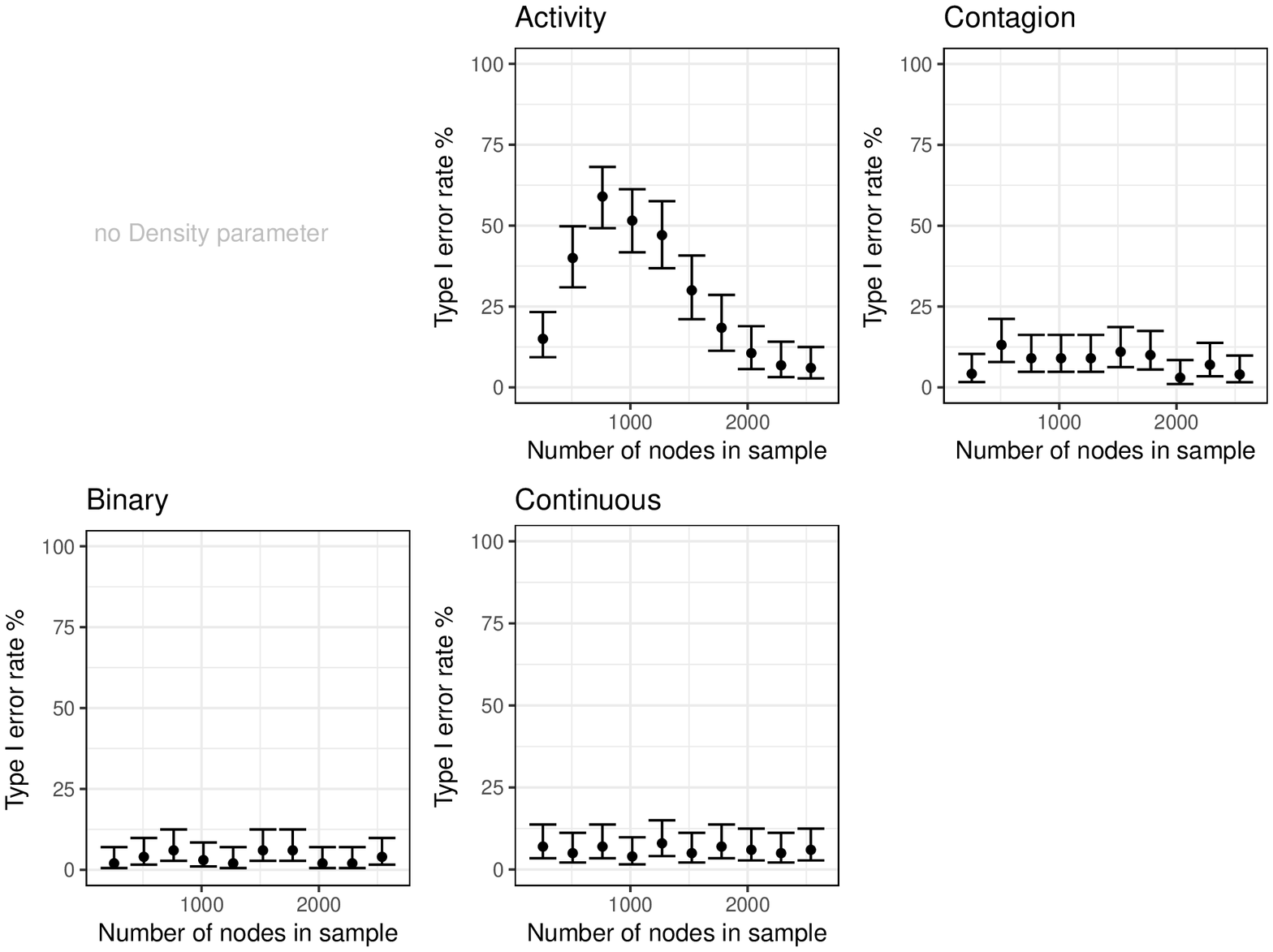}
  \caption{Effect on type I error rate of random node sampling. Add Health network $N=2539$.}
  \label{fig:addhealth_random_fpr}
\end{figure}

When using the empirical networks, however, the situation is not as
encouraging. Figure~\ref{fig:project90_random_fpr} shows the results
for the Project 90 network, and Figure~\ref{fig:addhealth_random_fpr}
for the Add Health network. For effects other than Activity, the type I error
rates are, sometimes, higher than desirable (particularly for
Contagion in the Project 90 network), but never much more than
25\%. However for the Activity parameter, the type I error rate can be
over 50\% (for the Add Health network), and the relationship between
sample size and type I error rate is noticeably not monotonic,
particularly for the Add Health network.

For both the Project 90 and Add Health networks, less than 1\% of the
ALAAM estimations just described do not converge.

Hence it seems that selecting nodes at random from an empirical
network (which may have, for example, a highly skewed degree
distribution), rather than following a network sampling scheme (such
as snowball sampling) can risk an unacceptably high (and particularly
difficult to estimate or predict, given the non-monotonicity in sample
size) type I error rate in some parameters.

\subsection{Simulation Study 2: Snowball Sampling}

The second study investigates the effect on ALAAM estimation of
network samples obtained via snowball sampling. Snowball samples are
obtained using 1, 2, or 3 waves. For the 2- and 3-wave snowball
sample, the number of seed nodes is varied from 1 to 20. For the
1-wave snowball sample, the number of seed nodes is varied from 1 to
100, since when there is only a single wave, many fewer nodes are
obtained in the sample, so a larger number of seed nodes may be
required. For each of these conditions, we investigate the effect of a
fixed-choice snowball sample, in which only up to $m$ ties are
followed (that is, degree censoring at $m$), with $m=3$ or $m=5$, as
well as the case where there is no degree censoring 
($m =\mbox{Inf}$).

\begin{figure}
  \includegraphics[width=\textwidth]{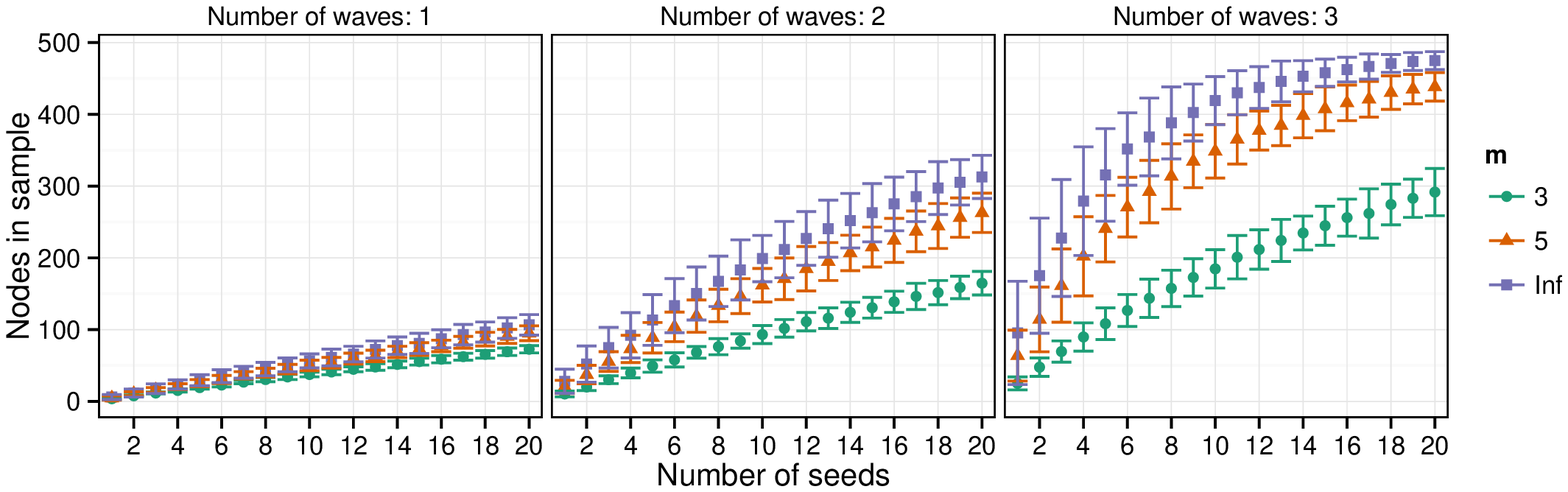}
  \caption{Size of the snowball sample with 1, 2, or 3 waves and degree censoring at $m=3$, $m=5$, or no degree censoring ($m=\mbox{Inf}$). Simulated ERGM network $N=500$.}
  \label{fig:samplesize}
\end{figure}

\begin{figure}
  \includegraphics[width=\textwidth]{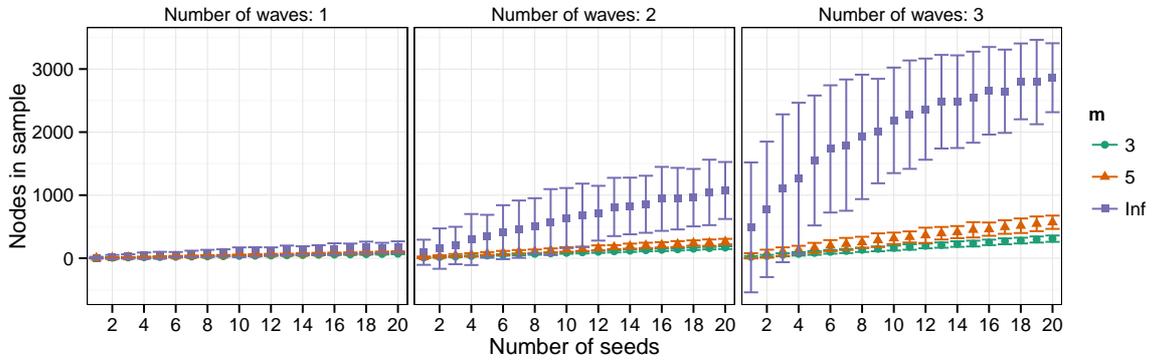}
  \caption{Size of the snowball sample with 1, 2, or 3 waves and degree censoring at $m=3$, $m=5$, or no degree censoring ($m=\mbox{Inf}$). Project 90 network $N=4430$.}
  \label{fig:project90_samplesize}
\end{figure}

\begin{figure}
  \includegraphics[width=\textwidth]{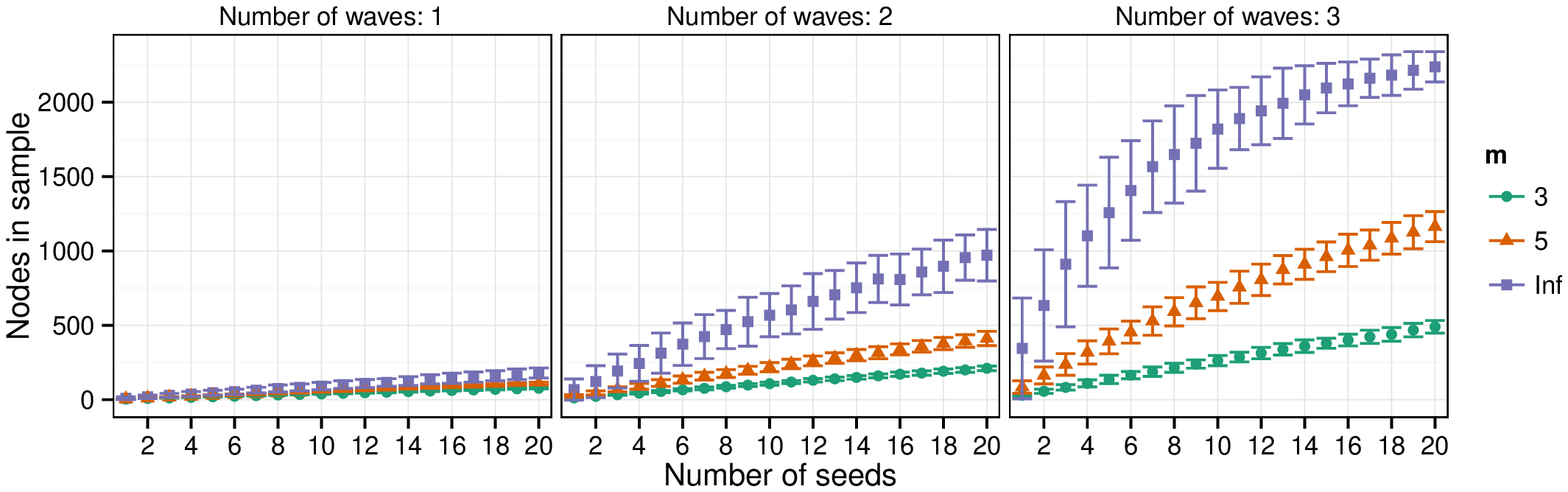}
  \caption{Size of the snowball sample with 1, 2, or 3 waves and degree censoring at $m=3$, $m=5$, or no degree censoring ($m=\mbox{Inf}$). Add Health network $N=2539$.}
  \label{fig:addhealth_samplesize}
\end{figure}

Figures~\ref{fig:samplesize}, \ref{fig:project90_samplesize} and
\ref{fig:addhealth_samplesize} show the size of the snowball samples
obtained with different snowball sampling parameters for the
simulated ERGM network, the Project 90 network, and the Add Health
network, respectively. Clearly, the sample sizes grows faster in the
number of seed nodes when more waves are used. Not using degree
censoring also results in large sample sizes, and also larger variance
in sample size.

\begin{figure}
  \includegraphics[width=\textwidth]{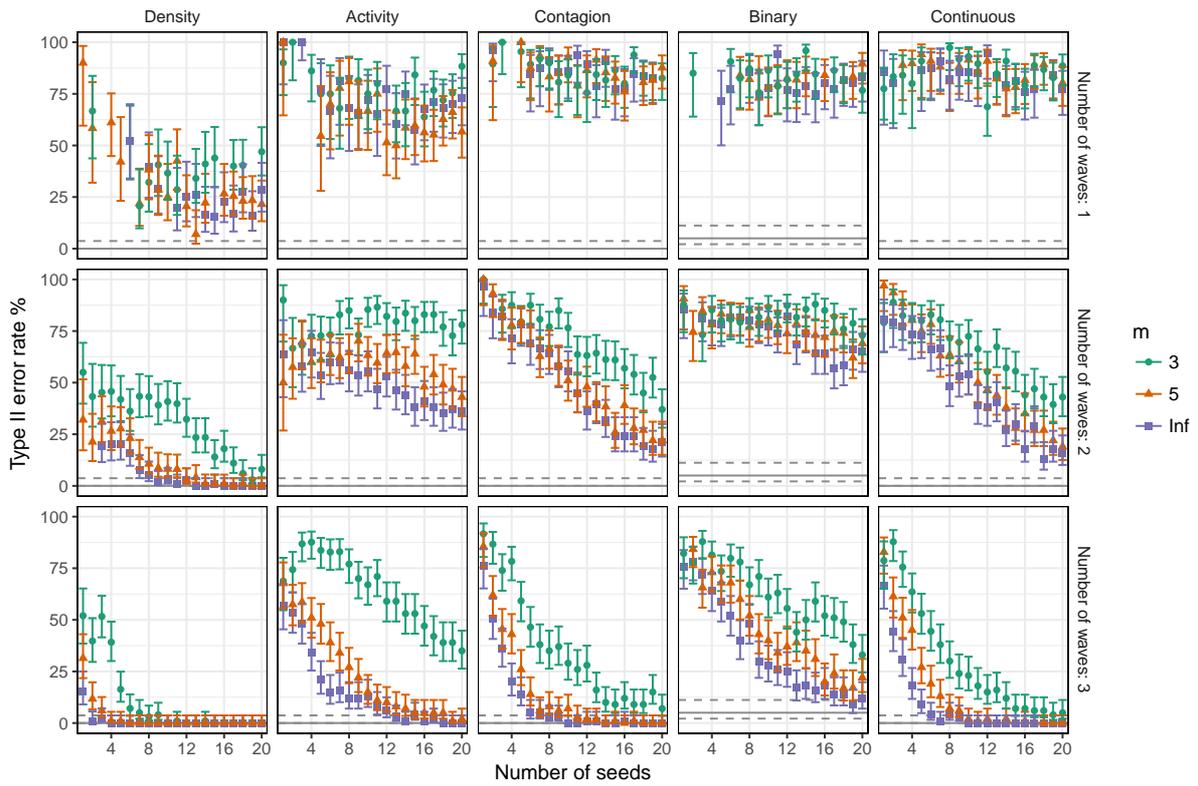}
  \caption{Effect on type II error rate of conditional estimation with snowball sampling with 1, 2, or 3 waves and degree censoring at $m=3$, $m=5$, or no degree censoring ($m=\mbox{Inf}$). Simulated ERGM network $N=500$. The solid horizontal line is the baseline error rate, with the dashed lines showing its 95\% confidence interval.}
  \label{fig:snowball_fnr}
\end{figure}

Figure~\ref{fig:snowball_fnr} shows the type II error rate, on network
samples from the simulated ERGM network, obtained by snowball sampling
with different numbers of waves and seeds, and with and without fixed
choice (at two different values) limiting the maximum number of ties
followed in the snowball sampling.  This shows that using more waves
in the snowball sampling gives higher power (lower type II error
rate). This is particularly noticeable for the Activity parameter,
which has extremely low power when using one or two waves, but can
achieve the same power as using the full network with 13 seeds (when
not using fixed choice) when using three waves.

It is clear that, for a given number of waves and seeds, there is
higher power with a higher value of $m$ (the number of links to
follow) in the fixed choice design, and higher still when not using a
fixed choice design (that is, following all links in the snowball
sampling).  For example, when using three waves, the Contagion parameter
has the same power as the baseline (using the entire network) with 10
or more seeds when not using fixed choice ($m=\mbox{Inf}$), but never
achieves this same power even with 20 seeds using fixed choice with
$m=3$.

\begin{figure}
  \includegraphics[width=\textwidth]{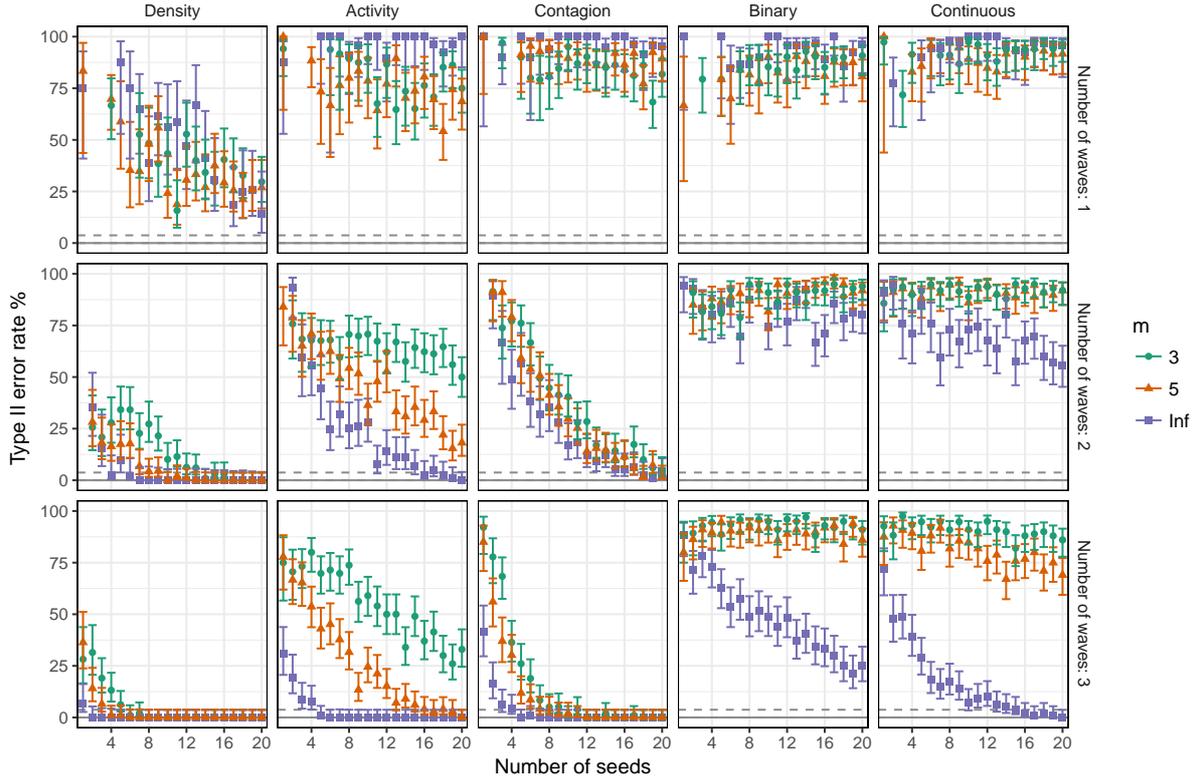}
  \caption{Effect on type II error rate of conditional estimation with snowball sampling with 1, 2, or 3 waves and degree censoring at $m=3$, $m=5$, or no degree censoring ($m=\mbox{Inf}$). Project 90 network ($N=4430$). The solid horizontal line is the baseline error rate, with the dashed lines showing its 95\% confidence interval.}
  \label{fig:project90_snowball_fnr}
\end{figure}

\begin{figure}
  \includegraphics[width=\textwidth]{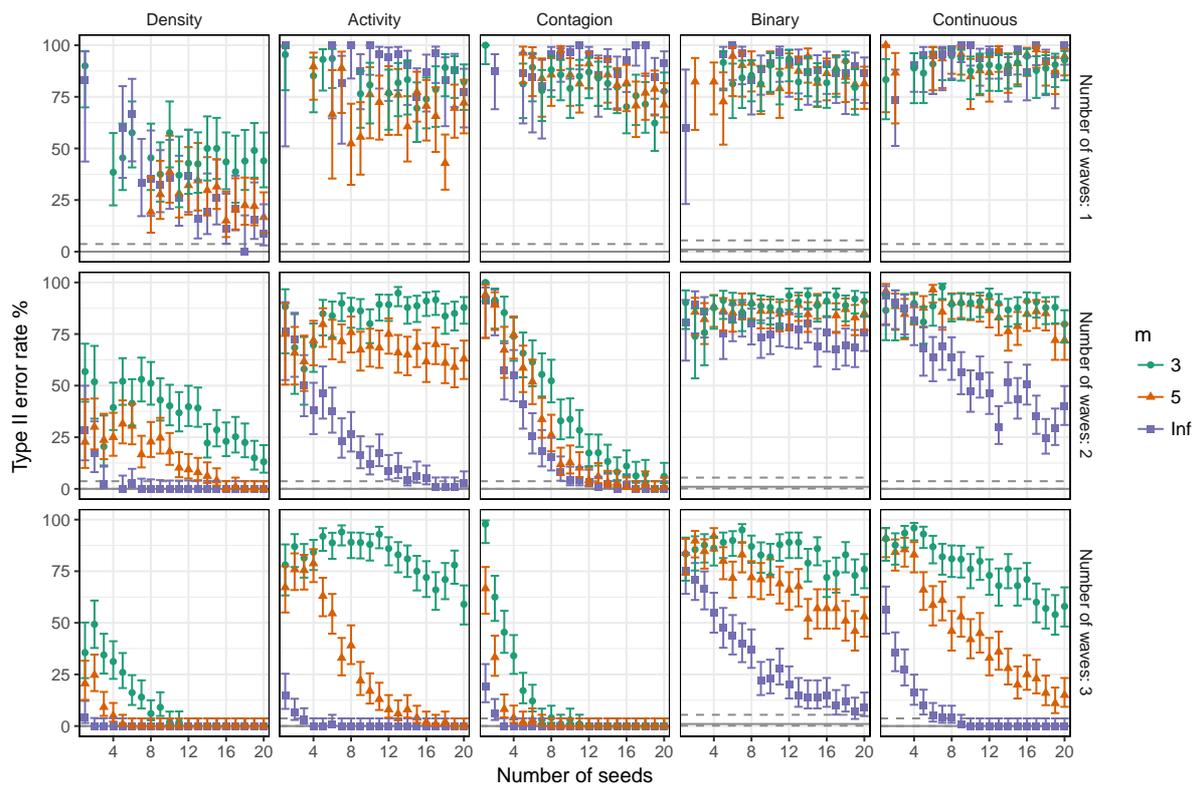}
  \caption{Effect on type II error rate of conditional estimation with snowball sampling with 1, 2, or 3 waves and degree censoring at $m=3$, $m=5$, or no degree censoring ($m=\mbox{Inf}$). Add Health network ($N=2539$). The solid horizontal line is the baseline error rate, with the dashed lines showing its 95\% confidence interval.}
  \label{fig:addhealth_snowball_fnr}
\end{figure}

The results are qualitatively similar, for the empirical networks,
Project 90 (Figure~\ref{fig:project90_snowball_fnr}) and Add Health
(Figure~\ref{fig:addhealth_snowball_fnr}). An exception is that the
advantage of not using degree censoring is even greater for the
empirical networks than in the simulated network (especially on the
Activity, Binary, and Continuous parameters), and that for the
Activity parameter on the Add Health network in particular, using a
fixed choice design with $m=3$ results in very low power even for 3
waves and 20 seeds.

\begin{figure}
  \includegraphics[width=\textwidth]{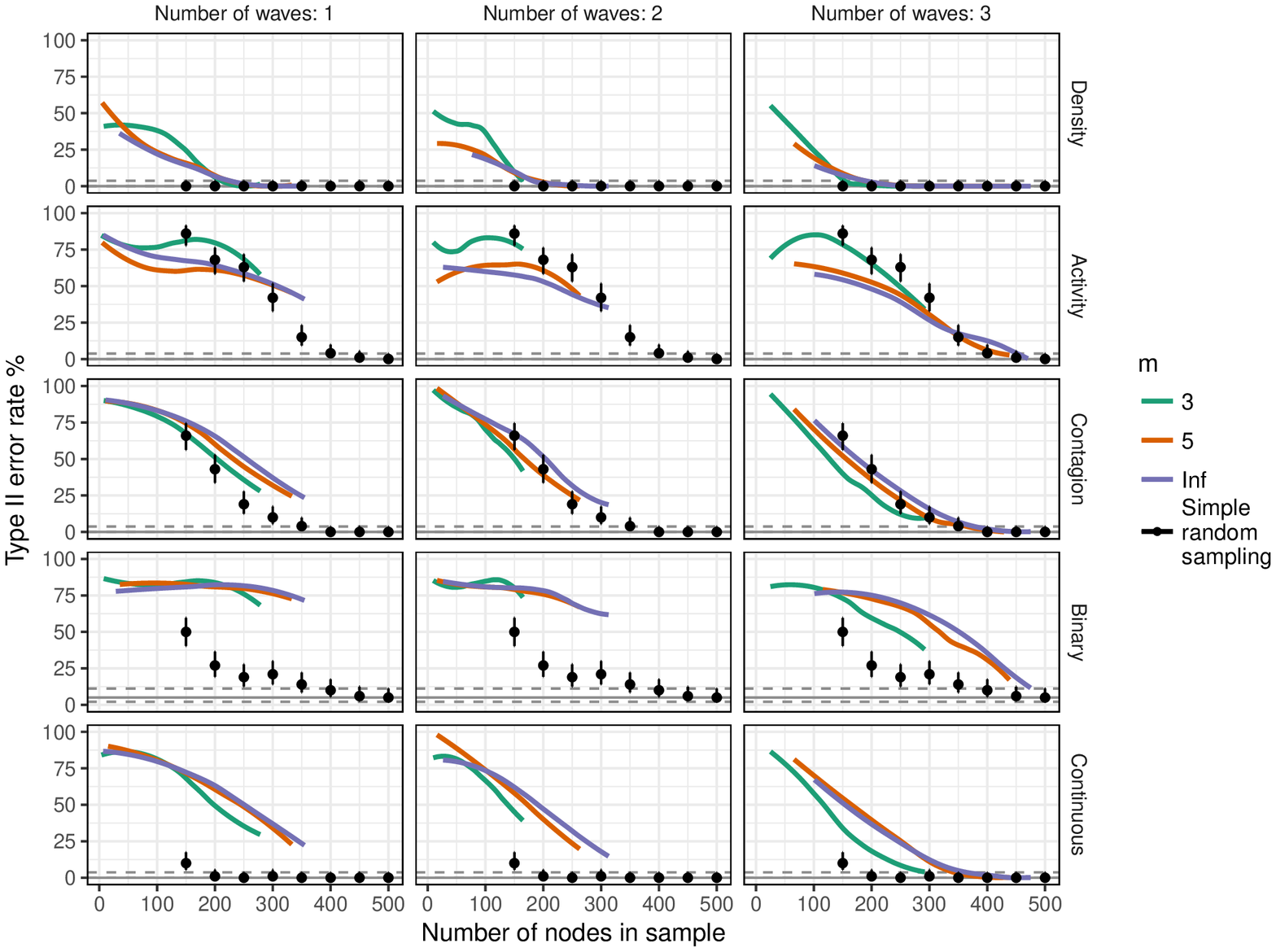}
  \caption{Type II error rate of conditional estimation plotted against sample size for snowball sampling with 1, 2, or 3 waves and degree censoring at $m=3$, $m=5$, or no degree censoring ($m=\mbox{Inf}$), as well as for simple random sampling. For snowball sampling, the number of seeds is varied from 1 to 100 for 1 wave, and 1 to 20 for 2 and 3 waves, and only the locally weighted polynomial regression (\unix{loess} in R) curve is shown, to make the figure clearer. Simulated ERGM network $N=500$. The solid horizontal line is the baseline error rate, with the dashed lines showing its 95\% confidence interval.}
\label{fig:snowball_fnr_size_smooth}  
\end{figure}

To address the question of whether these results are simply due to the
different sample sizes generated by the snowball sampling parameters
(waves, seeds, fixed choice size $m$), or if the sample structure
might be relevant, Figure~\ref{fig:snowball_fnr_size_smooth} shows the
type II error rate as a function of sample size, for both snowball
sampling and simple random sampling (that is, the results from
simulation study 1, described in the previous section).  These graphs
show that, for a given snowball sample size, there is (mostly) no
significant difference in the type II error rate for the different
snowball sampling parameters, except in some cases where fixed choice
$m=3$ has significantly different power from the others (\eg on
Activity on Binary with 3 waves). Snowball sampling has a lower type
II error rate than random sampling for the Activity parameter, and a
similar type II error rate for the Contagion parameter, when two or three
waves are used. However for the Binary and Continuous parameters,
snowball sampling has a significantly higher type II error rate than
random sampling over almost the entire range of sample sizes.

\begin{figure}
  \includegraphics[width=\textwidth]{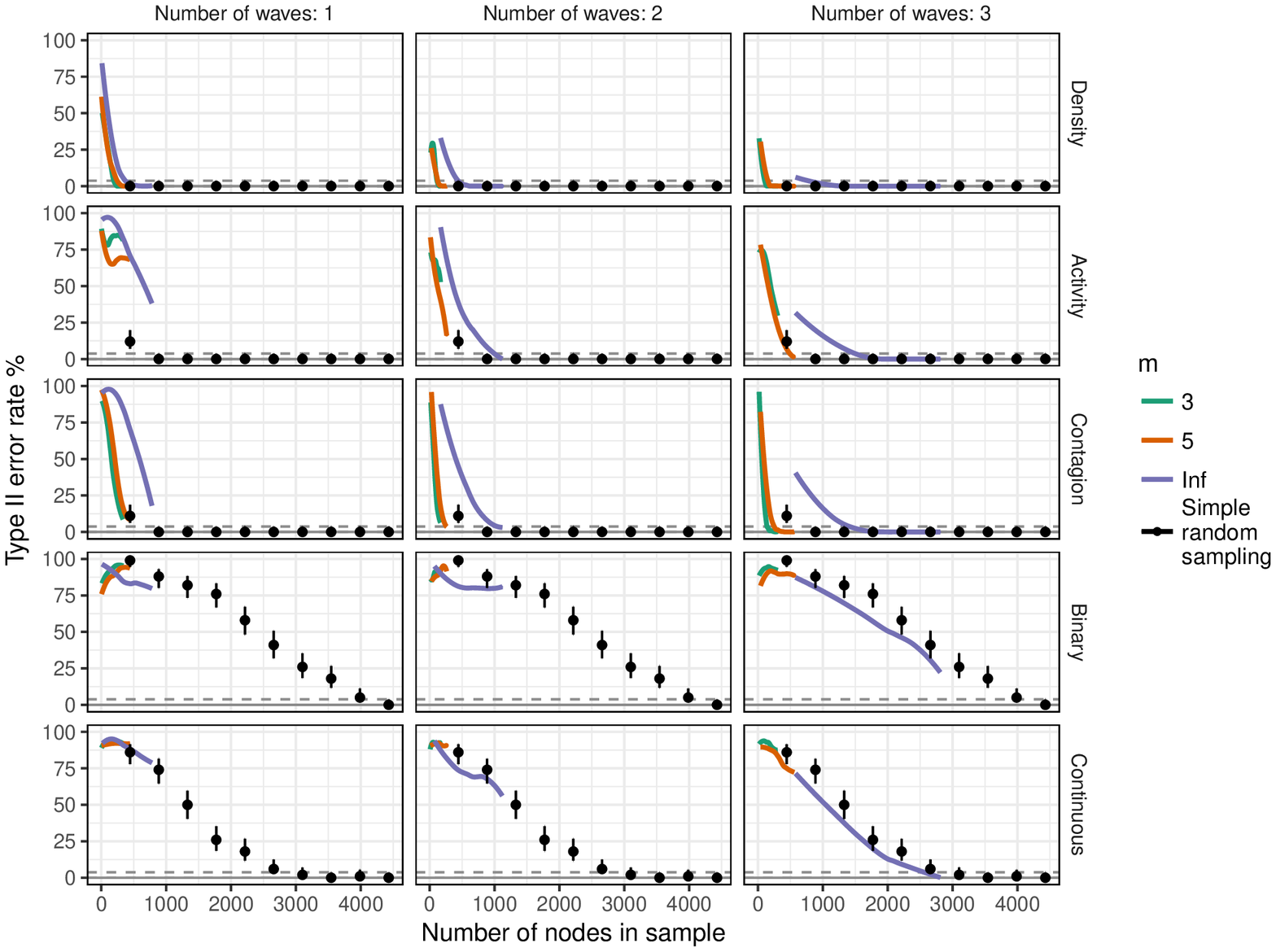}
  \caption{Type II error rate of conditional estimation plotted against sample size for snowball sampling with 1, 2, or 3 waves and degree censoring at $m=3$, $m=5$, or no degree censoring ($m=\mbox{Inf}$), as well as for simple random sampling. For snowball sampling, the number of seeds is varied from 1 to 100 for 1 wave, and 1 to 20 for 2 and 3 waves, and only the locally weighted polynomial regression (\unix{loess} in R) curve is shown, to make the figure clearer. Project 90 network ($N=4430$). The solid horizontal line is the baseline error rate, with the dashed lines showing its 95\% confidence interval.}
\label{fig:project90_snowball_fnr_size_smooth}  
\end{figure}

\begin{figure}
  \includegraphics[width=\textwidth]{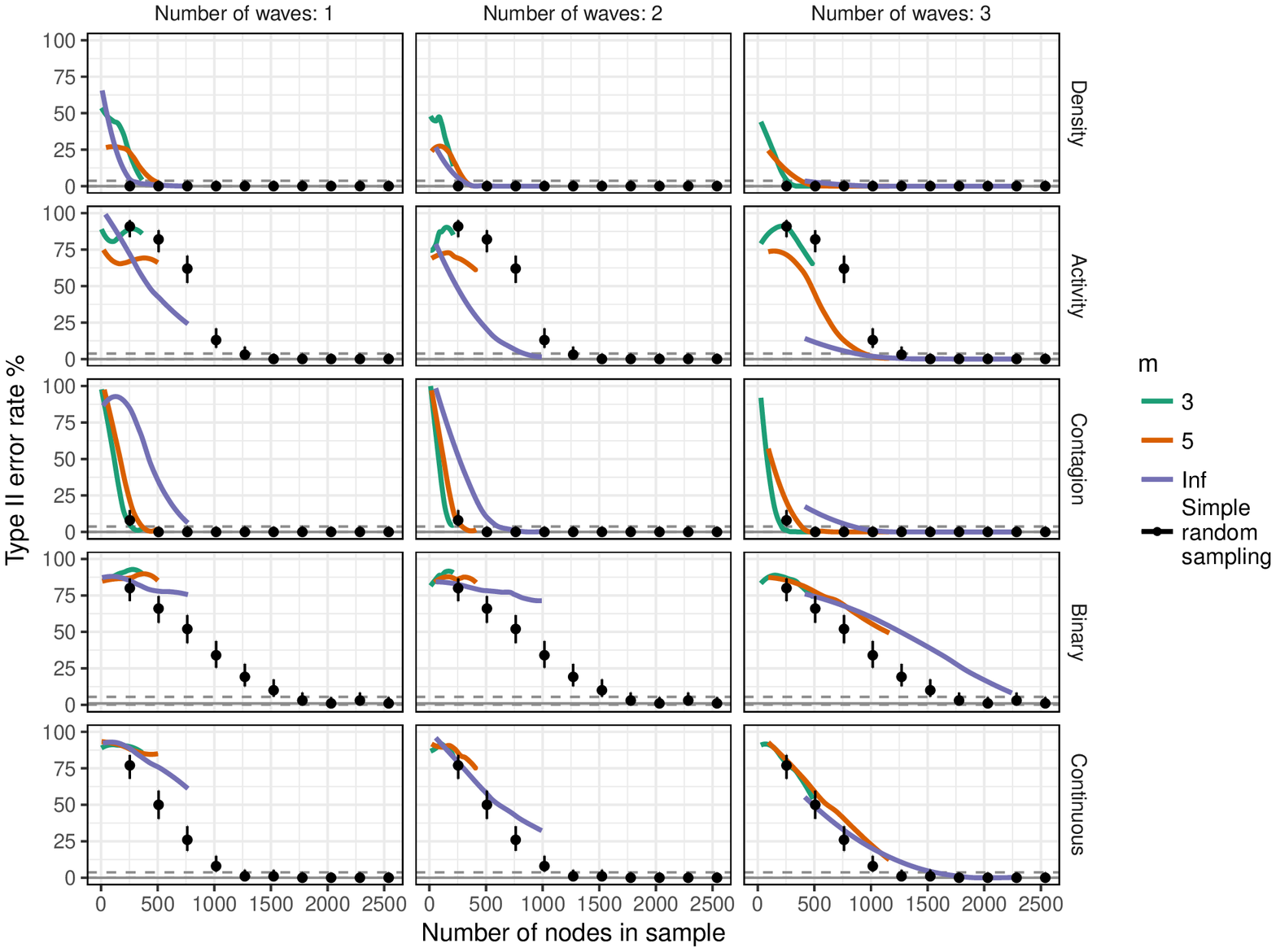}
  \caption{Type II error rate of conditional estimation plotted against sample size for snowball sampling with 1, 2, or 3 waves and degree censoring at $m=3$, $m=5$, or no degree censoring ($m=\mbox{Inf}$), as well as for simple random sampling. For snowball sampling, the number of seeds is varied from 1 to 100 for 1 wave, and 1 to 20 for 2 and 3 waves, and only the locally weighted polynomial regression (\unix{loess} in R) curve is shown, to make the figure clearer. Add Health network ($N=2539$). The solid horizontal line is the baseline error rate, with the dashed lines showing its 95\% confidence interval.}
\label{fig:addhealth_snowball_fnr_size_smooth}  
\end{figure}

Figures~\ref{fig:project90_snowball_fnr_size_smooth} and
\ref{fig:addhealth_snowball_fnr_size_smooth} show the corresponding results for
the Project 90 and Add Health networks, respectively. In general,
snowball sampling has a similar or lower type II error rate than
random node sampling for the same sample size. Exceptions to this are
for very small sample sizes, and for the Binary parameter in the Add
Health network.

\begin{figure}
  \includegraphics[width=\textwidth]{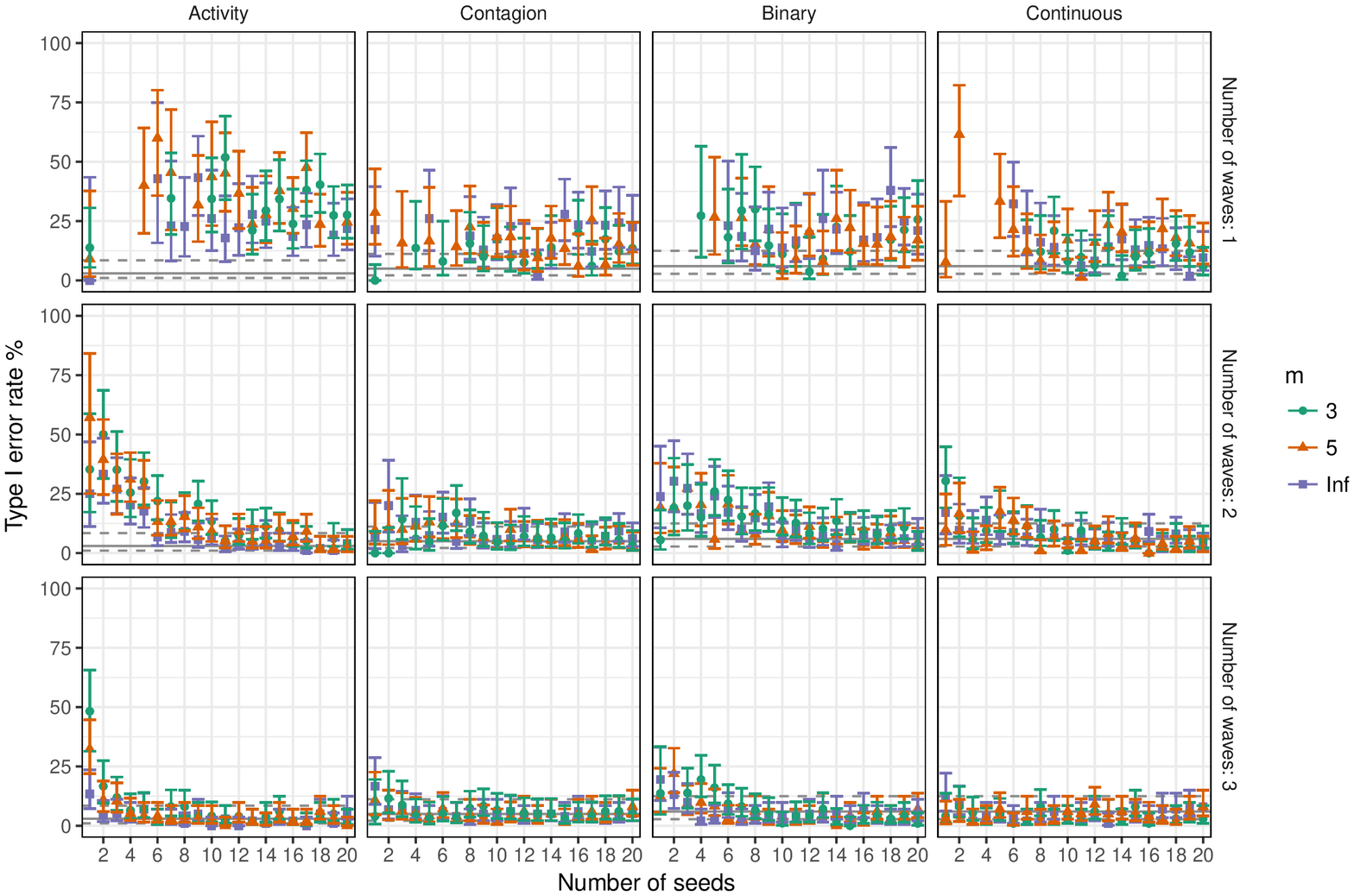}
  \caption{Effect on type I error rate of conditional estimation of snowball sampling with 1, 2, or 3 waves and degree censoring at $m=3$, $m=5$, or no degree censoring ($m=\mbox{Inf}$). Simulated ERGM network $N=500$. The solid horizontal line is the baseline error rate, with the dashed lines showing its 95\% confidence interval.}
  \label{fig:snowball_fpr}
\end{figure}

Figure~\ref{fig:snowball_fpr} shows that, in the simulated ERGM
network, when using three waves, except for very small numbers of
seeds, the type I error rate is not significantly different from the
baseline (using the entire network) on any of the parameters. However
when using fewer than three waves, the type I error rate can be far
higher.

\begin{figure}
  \includegraphics[width=\textwidth]{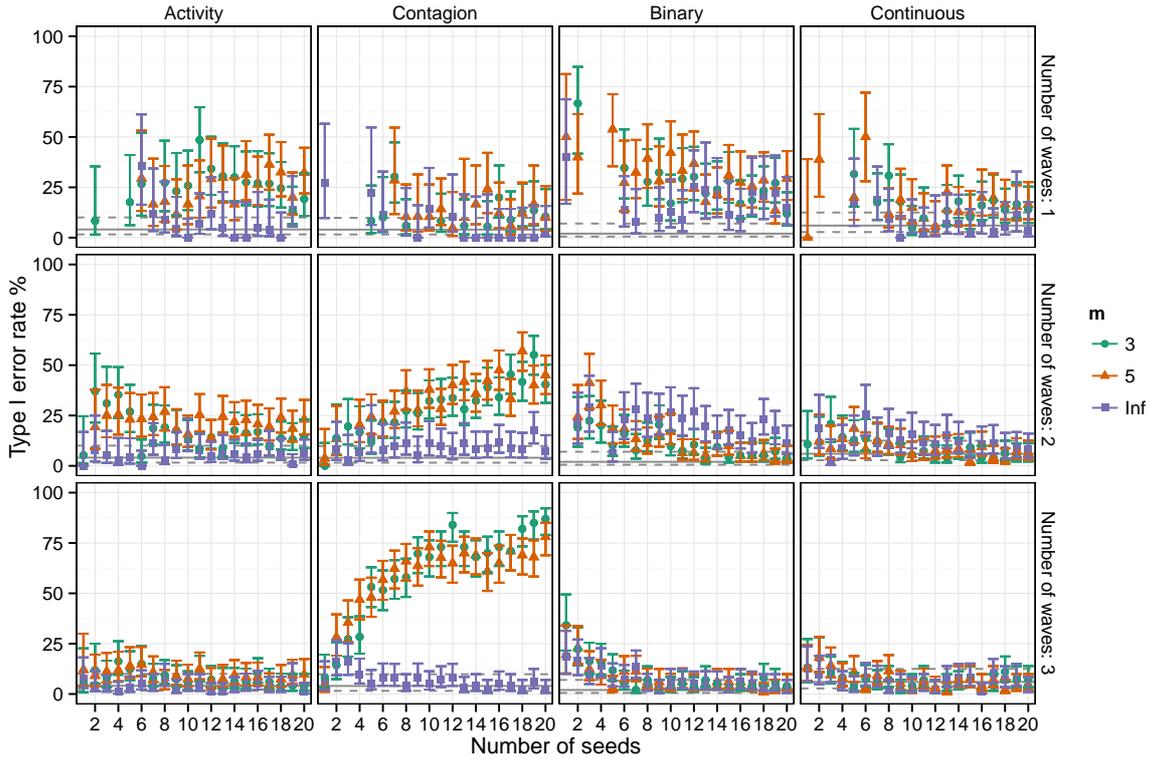}
  \caption{Effect on type I error rate of conditional estimation of snowball sampling with 1, 2, or 3 waves and degree censoring at $m=3$, $m=5$, or no degree censoring ($m=\mbox{Inf}$). Project 90 network ($N=4430$). The solid horizontal line is the baseline error rate, with the dashed lines showing its 95\% confidence interval.}
  \label{fig:project90_snowball_fpr}
\end{figure}

\begin{figure}
  \includegraphics[width=\textwidth]{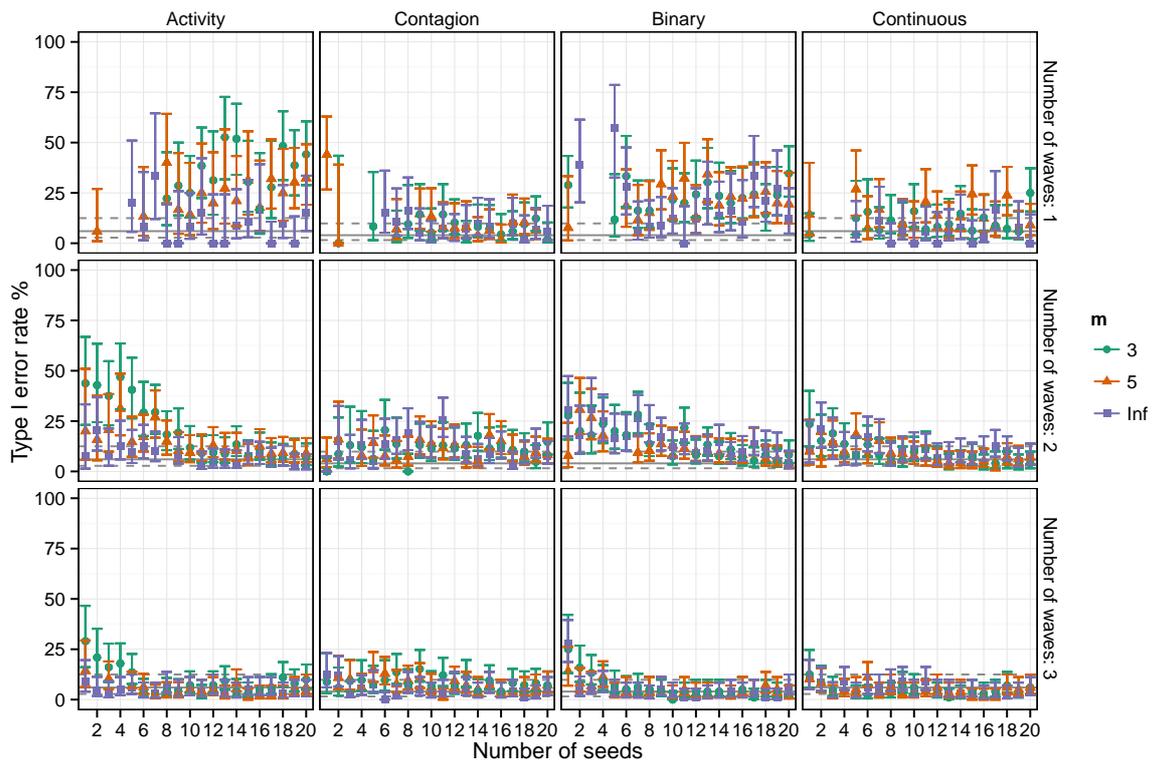}
  \caption{Effect on type I error rate of conditional estimation of snowball sampling with 1, 2, or 3 waves and degree censoring at $m=3$, $m=5$, or no degree censoring ($m=\mbox{Inf}$). Add Health network ($N=2539$). The solid horizontal line is the baseline error rate, with the dashed lines showing its 95\% confidence interval.}
  \label{fig:addhealth_snowball_fpr}
\end{figure}

Figures~\ref{fig:project90_snowball_fpr} and
\ref{fig:addhealth_snowball_fpr} show the corresponding results for
the Project 90 and Add Health networks, respectively. Similarly for
the results for the simulated ERGM network, using three waves results
in an acceptable type I error rate, however using fewer waves
frequently does not.  A very noticeable exception is in the Contagion
parameter in the Project 90 network
(Figure~\ref{fig:project90_snowball_fpr}) where the type I error rate
actually increases very rapidly to very high values when degree
censoring is used with three waves; however when degree censoring is
\emph{not} used, the type I error rate is not significantly different from
the baseline (the rate when the full network is used).

\begin{figure}
  \includegraphics[width=\textwidth]{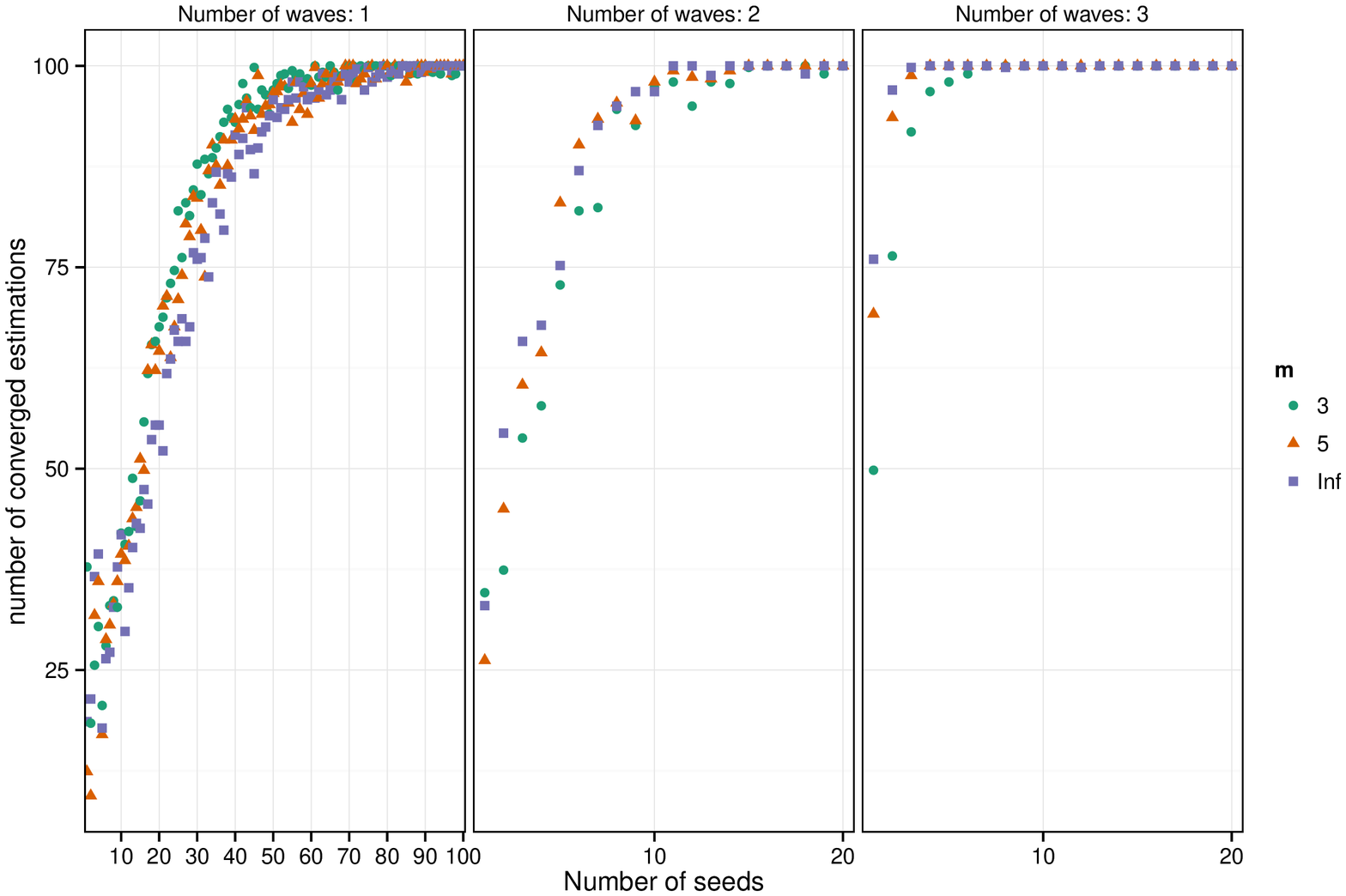}
  \caption{Number of converged estimates (out of 100) when using conditional estimation with snowball sampling with 1, 2, or 3 waves and degree censoring at $m=3$, $m=5$, or no degree censoring ($m=\mbox{Inf}$). Simulated ERGM network $N=500$.}
  \label{fig:converged}
\end{figure}

\begin{figure}
  \includegraphics[width=\textwidth]{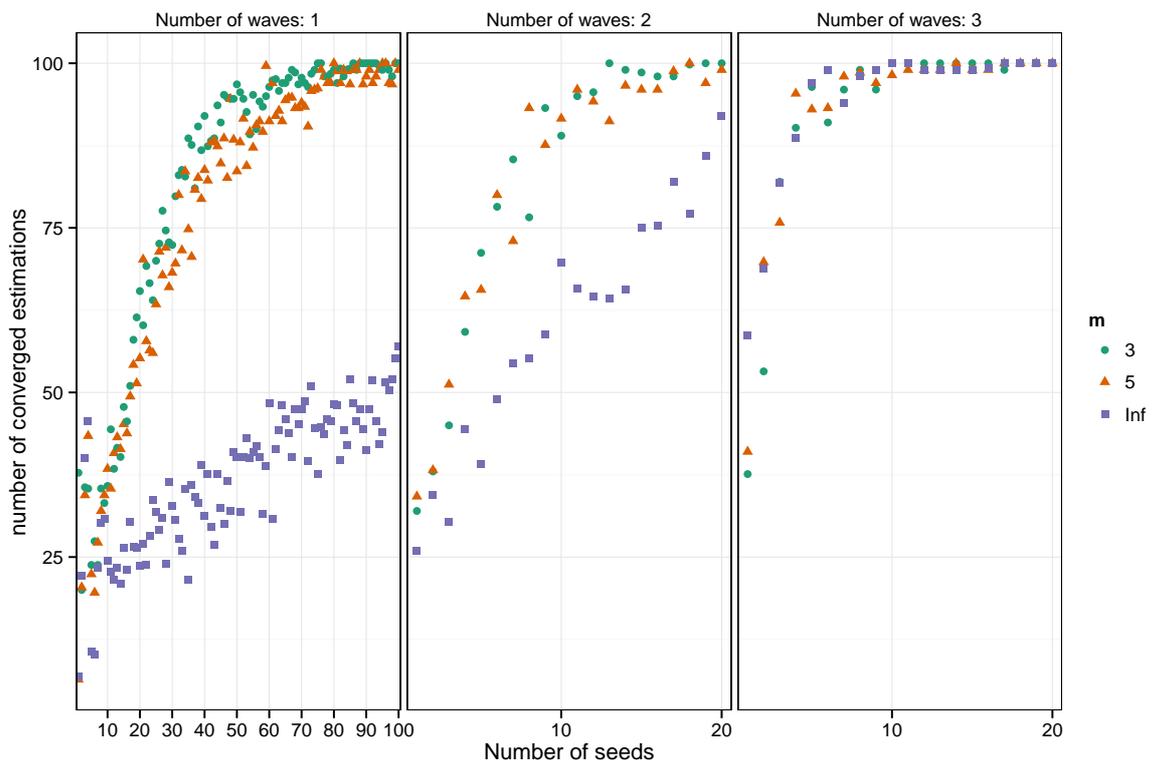}
  \caption{Number of converged estimates (out of 100) when using conditional estimation with snowball sampling with 1, 2, or 3 waves and degree censoring at $m=3$, $m=5$, or no degree censoring ($m=\mbox{Inf}$). Project 90 network ($N=4430$).}
  \label{fig:project90_converged}
\end{figure}

\begin{figure}
  \includegraphics[width=\textwidth]{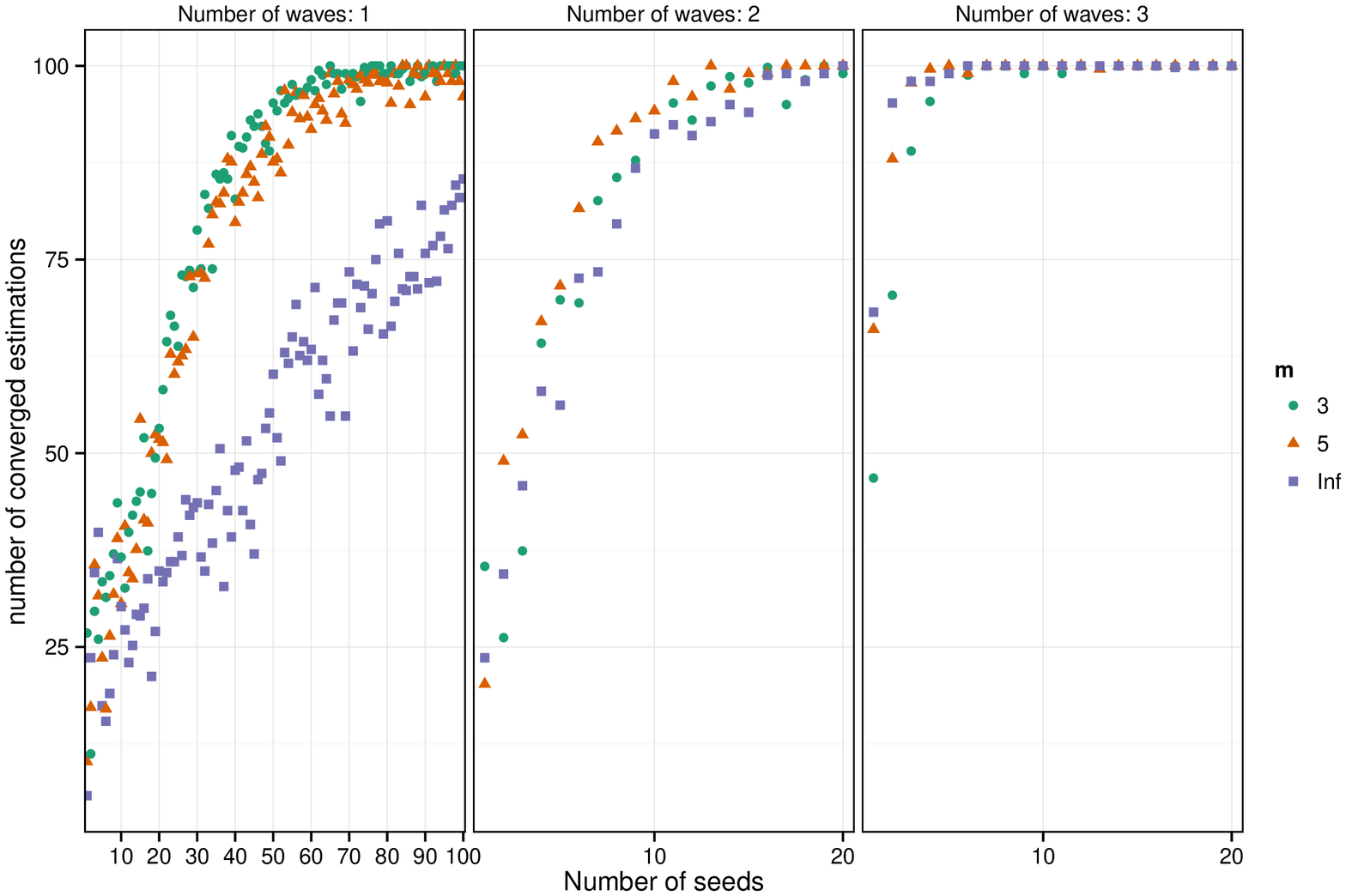}
  \caption{Number of converged estimates (out of 100) when using conditional estimation with snowball sampling with 1, 2, or 3 waves and degree censoring at $m=3$, $m=5$, or no degree censoring ($m=\mbox{Inf}$). Add Health network ($N=2539$).}
  \label{fig:addhealth_converged}
\end{figure}

Figures~\ref{fig:converged}, \ref{fig:project90_converged}, and
\ref{fig:addhealth_converged} show the number of converged estimates
for the simulated ERGM network, the Project 90 network, and the Add
Health network, respectively. Note that, unlike simulation study 1,
where nodes were included in the network sample at random, the network sample
is now a snowball sample, and conditional estimation is used to
estimate ALAAM parameters conditional on the snowball sampling
structure. As a result of this, it can be more difficult to obtain a
converged estimate, especially when a smaller number of waves are
used. In fact, when using fewer than three waves, and particularly when
using only one wave, the percentage of converged estimates can be very
low, unless a very large number of seeds is used. When using three
waves, however, most estimations converge even for a relatively small
number of seeds.

With this in mind, the type I error rate results
(Figures~\ref{fig:snowball_fpr}, \ref{fig:project90_snowball_fpr}, and
\ref{fig:addhealth_snowball_fpr}) are perhaps not as bad as they first
appear. Specifically, it would appear that in situations where the
type I error rate is likely to be unacceptably high, it is quite
likely that conditional estimation will not converge. Hence rather
than risking false positive conclusions, it is likely that in fact no
estimate could be obtained at all. (Whether this is actually a
preferable situation is perhaps arguable).

\section{Conclusion}
\label{sec:conclusion}

ALAAM parameter inference can work well even with a relatively small
sample taken from the original network, when the sample is obtained
by a snowball sample and the estimation is conditional on this sampling
structure.

Relatively small amounts of data missing at random from a network may not have
an adverse effect on ALAAM parameter inference, however the amount of
missing data, and the magnitude of its effect on error rates, may be
difficult to estimate. The main recommendation from this study is
that, rather than risk a possibly unknown amount of missing data, to use
a snowball sampling scheme to obtain a structured network sample
instead. Further, snowball sampling should use (at least) three
waves, and degree censoring should not be used (\ie all links should
be followed, rather than capping at an arbitrary number).

The main limitation of this study are that only three networks (one
simulated from an ERGM, and two empirical), were used, and that the
ALAAM model is relatively simple (it includes ``contagion'', probably
the most important and frequently used parameter for an ALAAM, but does
not include any triangular configurations, for example). Hence we
cannot generalize with any confidence about performance on other
networks, or more complex models, although the fact that the results
are similar on the simulated and empirical networks of different sizes
might be an encouraging sign that the results will at least be
applicable to networks ``like'' those tested. The methods described
here could be used on the actual network collected in empirical
studies as a sensitivity analysis of ALAAM parameter inference.

Another major limitation of this work is that, although two of the
networks used are empirical, in all cases both the attributes (nodal
covariates) were simulated, and the actual binary outcome variable was
simulated from an ALAAM. Hence, in some sense this is the easiest case
for estimation of an ALAAM, as it is known that the outcome was
generated from an ALAAM. Further work could be to validate ALAAM
estimation and conditional ALAAM estimation from snowball samples of
networks with empirical nodal covariates and outcome variable. Such
data is potentially available, for example, from the Project 90
network.

We have assumed, in the fixed choice snowball sampling design, that
the (up to) $m$ links followed from each node are chosen at random.
However in practice when conducting network data collection and asking
respondents to name up to $m$ friends, for example, the named contacts
are subject to cognitive biases \citep{freeman87} and hence not random.  
This could potentially lead to biases in the
sample that are not considered in this study.

This study is an early step in understanding how ALAAMs and
sociocentric network designs may be conducted effectively within
general community settings. In such settings, missing data, and/or the
necessity to obtain data on only a sample, not the full network, are
inevitable, but the research questions are often of paramount
importance. An interesting possibility raised by these findings is
whether long-established random sampling approaches could be
complemented with snowball and referral techniques for estimating
network-based determinants for individual outcomes. A wider program of
research is merited to determine this conclusively.

  \section*{Acknowledgments}
  
  This research was supported by a Victorian Life Sciences Computation
  Initiative (VLSCI) grant number VR0261 on its Peak Computing Facility
  at the University of Melbourne, an initiative of the Victorian
  Government, Australia.  We also used the University of Melbourne ITS
  High Performance Computing facilities.

  This research uses data from Add Health, a program project directed
  by Kathleen Mullan Harris and designed by J. Richard Udry, Peter
  S. Bearman, and Kathleen Mullan Harris at the University of North
  Carolina at Chapel Hill, and funded by grant P01-HD31921 from the
  Eunice Kennedy Shriver National Institute of Child Health and Human
  Development, with cooperative funding from 23 other federal agencies
  and foundations. Special acknowledgment is due Ronald R. Rindfuss
  and Barbara Entwisle for assistance in the original
  design. Information on how to obtain the Add Health data files is
  available on the Add Health website
  (\url{http://www.cpc.unc.edu/addhealth}). No direct support was received
  from grant P01-HD31921 for this analysis.

\bibliographystyle{abbrvnat}

\begin{thebibliography}{75}
\providecommand{\natexlab}[1]{#1}
\providecommand{\url}[1]{\texttt{#1}}
\expandafter\ifx\csname urlstyle\endcsname\relax
  \providecommand{\doi}[1]{doi: #1}\else
  \providecommand{\doi}{doi: \begingroup \urlstyle{rm}\Url}\fi

\bibitem[Anselin(1990)]{anselin90}
L.~Anselin.
\newblock Some robust approaches to testing and estimation in spatial
  econometrics.
\newblock \emph{Regional Science and Urban Economics}, 20\penalty0
  (2):\penalty0 141--163, 1990.

\bibitem[Borgatti et~al.(2006)Borgatti, Carley, and Krackhardt]{borgatti06b}
S.~P. Borgatti, K.~M. Carley, and D.~Krackhardt.
\newblock On the robustness of centrality measures under conditions of
  imperfect data.
\newblock \emph{Social Networks}, 28\penalty0 (2):\penalty0 124--136, 2006.

\bibitem[Bryant et~al.(2014)Bryant, Waters, Gibbs, Gallagher, Pattison, Lusher,
  MacDougall, Harms, Block, Snowdon, Sinnott, Ireton, Richardson, and
  Forbes]{bryant14}
R.~A. Bryant, E.~Waters, L.~Gibbs, H.~C. Gallagher, P.~Pattison, D.~Lusher,
  C.~MacDougall, L.~Harms, K.~Block, E.~Snowdon, V.~Sinnott, G.~Ireton,
  J.~Richardson, and D.~Forbes.
\newblock Psychological outcomes following the {Victorian Black Saturday}
  bushfires.
\newblock \emph{Australian and New Zealand Journal of Psychiatry}, 48\penalty0
  (7):\penalty0 634--643, 2014.

\bibitem[Cliff and Ord(1981)]{cliff81}
A.~D. Cliff and J.~K. Ord.
\newblock \emph{Spatial processes: models \& applications}.
\newblock Taylor \& Francis, 1981.

\bibitem[Coleman(1958)]{coleman58}
J.~S. Coleman.
\newblock Relational analysis: the study of social organizations with survey
  methods.
\newblock \emph{Human Organization}, 17\penalty0 (4):\penalty0 28--36, 1958.

\bibitem[Cs\'ardi and Nepusz(2006)]{csardi06}
G.~Cs\'ardi and T.~Nepusz.
\newblock The igraph software package for complex network research.
\newblock \emph{InterJournal Complex Systems}, 1695, 2006.
\newblock URL \url{http://igraph.sf.net}.

\bibitem[Daraganova(2009)]{daraganova09}
G.~Daraganova.
\newblock \emph{Statistical models for social networks and network-mediated
  social influence processes: Theory and application}.
\newblock PhD thesis, The University of Melbourne, 2009.

\bibitem[Daraganova and Pattison(2013)]{daraganova13b}
G.~Daraganova and P.~Pattison.
\newblock Autologistic actor attribute model analysis of unemployment: dual
  importance of who you know and where you live.
\newblock In D.~Lusher, J.~Koskinen, and G.~Robins, editors, \emph{Exponential
  Random Graph Models for Social Networks}, chapter~18, pages 237--247.
  Cambridge University Press, New York, 2013.

\bibitem[Daraganova and Robins(2013)]{daraganova13}
G.~Daraganova and G.~Robins.
\newblock Autologistic actor attribute models.
\newblock In D.~Lusher, J.~Koskinen, and G.~Robins, editors, \emph{Exponential
  Random Graph Models for Social Networks}, chapter~9, pages 102--114.
  Cambridge University Press, New York, 2013.

\bibitem[Davison and Hinkley(1997)]{davison97}
A.~C. Davison and D.~V. Hinkley.
\newblock \emph{Bootstrap Methods and Their Application}.
\newblock Cambridge University Press, Cambridge, 1997.

\bibitem[De~Silva et~al.(2005)De~Silva, McKenzie, Harpham, and
  Huttly]{desilva05}
M.~J. De~Silva, K.~McKenzie, T.~Harpham, and S.~R. Huttly.
\newblock Social capital and mental illness: a systematic review.
\newblock \emph{Journal of Epidemiology and Community Health}, 59\penalty0
  (8):\penalty0 619--627, 2005.

\bibitem[Dekker et~al.(2007)Dekker, Krackhardt, and Snijders]{dekker07}
D.~Dekker, D.~Krackhardt, and T.~A. Snijders.
\newblock Sensitivity of {MRQAP} tests to collinearity and autocorrelation
  conditions.
\newblock \emph{Psychometrika}, 72\penalty0 (4):\penalty0 563--581, 2007.

\bibitem[Dirkzwager et~al.(2006)Dirkzwager, Grievink, Van~der Velden, and
  Yzermans]{dirkzwager06}
A.~J. Dirkzwager, L.~Grievink, P.~G. Van~der Velden, and C.~J. Yzermans.
\newblock Risk factors for psychological and physical health problems after a
  man-made disaster.
\newblock \emph{The British Journal of Psychiatry}, 189\penalty0 (2):\penalty0
  144--149, 2006.

\bibitem[Dittrich et~al.(2017)Dittrich, Leenders, and Mulder]{dittrich17}
D.~Dittrich, R.~T.~A. Leenders, and J.~Mulder.
\newblock Bayesian estimation of the network autocorrelation model.
\newblock \emph{Social Networks}, 48:\penalty0 213--236, 2017.

\bibitem[Dittrich et~al.(2019)Dittrich, Leenders, and Mulder]{dittrich19}
D.~Dittrich, R.~T.~A. Leenders, and J.~Mulder.
\newblock Network autocorrelation modeling: A {Bayes} factor approach for
  testing (multiple) precise and interval hypotheses.
\newblock \emph{Sociological Methods \& Research}, 48\penalty0 (3):\penalty0
  642--676, 2019.

\bibitem[Doreian(1981)]{doreian81}
P.~Doreian.
\newblock Estimating linear models with spatially distributed data.
\newblock \emph{Sociological Methodology}, 12:\penalty0 359--388, 1981.

\bibitem[Efron(1987)]{efron87}
B.~Efron.
\newblock Better bootstrap confidence intervals.
\newblock \emph{Journal of the American Statistical Association}, 82\penalty0
  (397):\penalty0 171--185, 1987.

\bibitem[Frank and Strauss(1986)]{frank86}
O.~Frank and D.~Strauss.
\newblock Markov graphs.
\newblock \emph{Journal of the American Statistical Association}, 81\penalty0
  (395):\penalty0 832--842, 1986.

\bibitem[Freeman et~al.(1987)Freeman, Romney, and Freeman]{freeman87}
L.~C. Freeman, A.~K. Romney, and S.~C. Freeman.
\newblock Cognitive structure and informant accuracy.
\newblock \emph{American Anthropologist}, 89\penalty0 (2):\penalty0 310--325,
  1987.

\bibitem[Friedkin(1990)]{friedkin90}
N.~E. Friedkin.
\newblock Social networks in structural equation models.
\newblock \emph{Social Psychology Quarterly}, pages 316--328, 1990.

\bibitem[Gibbs et~al.(2013)Gibbs, Waters, Bryant, Pattison, Lusher, Harms,
  Richardson, MacDougall, Block, Snowdon, et~al.]{gibbs13}
L.~Gibbs, E.~Waters, R.~A. Bryant, P.~Pattison, D.~Lusher, L.~Harms,
  J.~Richardson, C.~MacDougall, K.~Block, E.~Snowdon, et~al.
\newblock Beyond bushfires: Community, resilience and recovery --- a
  longitudinal mixed method study of the medium to long term impacts of
  bushfires on mental health and social connectedness.
\newblock \emph{BMC Public Health}, 13\penalty0 (1):\penalty0 1036, 2013.

\bibitem[Gibbs et~al.(2015)Gibbs, Howell-Meurs, Block, Lusher, Richardson,
  MacDougall, Waters, Harms, et~al.]{gibbs15}
L.~Gibbs, S.~Howell-Meurs, K.~Block, D.~Lusher, J.~Richardson, C.~MacDougall,
  E.~Waters, L.~Harms, et~al.
\newblock Community wellbeing: applications for a disaster context.
\newblock \emph{Australian Journal of Emergency Management}, 30:\penalty0
  20--24, 2015.

\bibitem[Goel and Salganik(2010)]{goel10}
S.~Goel and M.~J. Salganik.
\newblock Assessing respondent-driven sampling.
\newblock \emph{Proceedings of the National Academy of Sciences of the USA},
  107\penalty0 (15):\penalty0 6743--6747, 2010.

\bibitem[Goodman(1961)]{goodman61}
L.~A. Goodman.
\newblock Snowball sampling.
\newblock \emph{The Annals of Mathematical Statistics}, pages 148--170, 1961.

\bibitem[Goodman(2011)]{goodman11}
L.~A. Goodman.
\newblock Comment: On respondent-driven sampling and snowball sampling in
  hard-to-reach populations and snowball sampling not in hard-to-reach
  populations.
\newblock \emph{Sociological Methodology}, 41\penalty0 (1):\penalty0 347--353,
  2011.

\bibitem[Handcock and Gile(2010)]{handcock10}
M.~S. Handcock and K.~J. Gile.
\newblock Modeling social networks from sampled data.
\newblock \emph{The Annals of Applied Statistics}, 4\penalty0 (1):\penalty0
  5--25, 2010.

\bibitem[Handcock and Gile(2011)]{handcock11}
M.~S. Handcock and K.~J. Gile.
\newblock Comment: On the concept of snowball sampling.
\newblock \emph{Sociological Methodology}, 41\penalty0 (1):\penalty0 367--371,
  2011.

\bibitem[Harris and Udry(2015)]{addhealth}
K.~M. Harris and R.~J. Udry.
\newblock {National Longitudinal Study of Adolescent to Adult Health (Add
  Health) Wave I, 1994-1995}, 2015.
\newblock URL \url{https://doi.org/10.15139/S3/11900}.

\bibitem[Heckathorn(2011)]{heckathorn11}
D.~D. Heckathorn.
\newblock Comment: Snowball versus respondent-driven sampling.
\newblock \emph{Sociological Methodology}, 41\penalty0 (1):\penalty0 355--366,
  2011.

\bibitem[Holland and Leinhardt(1973)]{holland73}
P.~W. Holland and S.~Leinhardt.
\newblock The structural implications of measurement error in sociometry†.
\newblock \emph{Journal of Mathematical Sociology}, 3\penalty0 (1):\penalty0
  85--111, 1973.

\bibitem[Illenberger and Fl\"{o}tter\"{o}d(2012)]{illenberger12}
J.~Illenberger and G.~Fl\"{o}tter\"{o}d.
\newblock Estimating network properties from snowball sampled data.
\newblock \emph{Social Networks}, 34\penalty0 (4):\penalty0 701--711, 2012.

\bibitem[Kashima et~al.(2013)Kashima, Wilson, Lusher, Pearson, and
  Pearson]{kashima13}
Y.~Kashima, S.~Wilson, D.~Lusher, L.~J. Pearson, and C.~Pearson.
\newblock The acquisition of perceived descriptive norms as social category
  learning in social networks.
\newblock \emph{Social Networks}, 35\penalty0 (4):\penalty0 711--719, 2013.

\bibitem[Kawachi et~al.(2013)Kawachi, Takao, and Subramanian]{kawachi13}
I.~Kawachi, S.~Takao, and S.~V. Subramanian.
\newblock \emph{Global perspectives on social capital and health}.
\newblock Springer, New York, 2013.

\bibitem[Kessler et~al.(2002)Kessler, Andrews, Colpe, Mroczek, Normand,
  Walters, and Zaslavsky]{kessler02}
R.~G. Kessler, G.~Andrews, L.~J. Colpe, D.~K. Mroczek, S.-L.~T. Normand, E.~E.
  Walters, and A.~M. Zaslavsky.
\newblock Short screening scales to monitor population prevalences and trends
  in non-specific psychological distress.
\newblock \emph{Psychological Medicine}, 32:\penalty0 959--976, 2002.

\bibitem[Kiuru et~al.(2012)Kiuru, Burk, Laursen, Nurmi, and
  Salmela-Aro]{kiuru12}
N.~Kiuru, W.~J. Burk, B.~Laursen, J.-E. Nurmi, and K.~Salmela-Aro.
\newblock Is depression contagious? a test of alternative peer socialization
  mechanisms of depressive symptoms in adolescent peer networks.
\newblock \emph{Journal of Adolescent Health}, 50\penalty0 (3):\penalty0
  250--255, 2012.

\bibitem[Klovdahl et~al.(1994)Klovdahl, Potterat, Woodhouse, Muth, Muth, and
  Darrow]{klovdahl94}
A.~S. Klovdahl, J.~J. Potterat, D.~E. Woodhouse, J.~B. Muth, S.~Q. Muth, and
  W.~W. Darrow.
\newblock Social networks and infectious disease: The {Colorado Springs} study.
\newblock \emph{Social Science \& Medicine}, 38\penalty0 (1):\penalty0 79--88,
  1994.

\bibitem[Koskinen et~al.(2013)Koskinen, Robins, Wang, and
  Pattison]{koskinen13a}
J.~H. Koskinen, G.~L. Robins, P.~Wang, and P.~E. Pattison.
\newblock Bayesian analysis for partially observed network data, missing ties,
  attributes and actors.
\newblock \emph{Social Networks}, 35\penalty0 (4):\penalty0 514--527, 2013.

\bibitem[Kossinets(2006)]{kossinets06}
G.~Kossinets.
\newblock Effects of missing data in social networks.
\newblock \emph{Social Networks}, 28\penalty0 (3):\penalty0 247--268, 2006.

\bibitem[Krackhardt(1988)]{krackhardt88b}
D.~Krackhardt.
\newblock Predicting with networks: Nonparametric multiple regression analysis
  of dyadic data.
\newblock \emph{Social Networks}, 10\penalty0 (4):\penalty0 359--381, 1988.

\bibitem[Kurant et~al.(2011)Kurant, Markopoulou, and Thiran]{kurant11}
M.~Kurant, A.~Markopoulou, and P.~Thiran.
\newblock Towards unbiased {BFS} sampling.
\newblock \emph{IEEE Journal on Selected Areas in Communications}, 29\penalty0
  (9):\penalty0 1799--1809, 2011.

\bibitem[Lee et~al.(2006)Lee, Kim, and Jeong]{lee06}
S.~H. Lee, P.-J. Kim, and H.~Jeong.
\newblock Statistical properties of sampled networks.
\newblock \emph{Physical Review E}, 73:\penalty0 016102, 2006.

\bibitem[Leenders(2002)]{leenders02}
R.~T.~A. Leenders.
\newblock Modeling social influence through network autocorrelation:
  constructing the weight matrix.
\newblock \emph{Social networks}, 24\penalty0 (1):\penalty0 21--47, 2002.

\bibitem[Letina(2016)]{letina16}
S.~Letina.
\newblock Network and actor attribute effects on the performance of researchers
  in two fields of social science in a small peripheral community.
\newblock \emph{Journal of Informetrics}, 10\penalty0 (2):\penalty0 571--595,
  2016.

\bibitem[Letina et~al.(2016)Letina, Robins, and
  Masli{\'c}~Ser{\v{s}}i{\'c}]{letina16b}
S.~Letina, G.~Robins, and D.~Masli{\'c}~Ser{\v{s}}i{\'c}.
\newblock Reaching out from a small scientific community: the social influence
  models of collaboration across national and disciplinary boundaries for
  scientists in three fields of social sciences.
\newblock \emph{Revija za sociologiju}, 46\penalty0 (2):\penalty0 103--139,
  2016.

\bibitem[Lusher et~al.(2013)Lusher, Koskinen, and Robins]{lusher13}
D.~Lusher, J.~Koskinen, and G.~Robins, editors.
\newblock \emph{Exponential Random Graph Models for Social Networks}.
\newblock Structural Analysis in the Social Sciences. Cambridge University
  Press, New York, 2013.

\bibitem[Mizruchi and Neuman(2008)]{mizruchi08}
M.~S. Mizruchi and E.~J. Neuman.
\newblock The effect of density on the level of bias in the network
  autocorrelation model.
\newblock \emph{Social Networks}, 30\penalty0 (3):\penalty0 190--200, 2008.

\bibitem[Moody(2001)]{moody01}
J.~Moody.
\newblock Peer influence groups: Identifying dense clusters in large networks.
\newblock \emph{Social Networks}, 23\penalty0 (4):\penalty0 261--283, 2001.

\bibitem[Neuman and Mizruchi(2010)]{neuman10}
E.~J. Neuman and M.~S. Mizruchi.
\newblock Structure and bias in the network autocorrelation model.
\newblock \emph{Social Networks}, 32\penalty0 (4):\penalty0 290--300, 2010.

\bibitem[Newman(2003)]{newman03}
M.~E.~J. Newman.
\newblock Ego-centered netwoks and the ripple effect.
\newblock \emph{Social Networks}, 25\penalty0 (1):\penalty0 83--95, 2003.

\bibitem[Newman(2006)]{newman06}
M.~E.~J. Newman.
\newblock Finding community structure in networks using the eigenvectors of
  matrices.
\newblock \emph{Physical Review E}, 74\penalty0 (3):\penalty0 036104, 2006.

\bibitem[Norris et~al.(2008)Norris, Stevens, Pfefferbaum, Wyche, and
  Pfefferbaum]{norris08}
F.~H. Norris, S.~P. Stevens, B.~Pfefferbaum, K.~F. Wyche, and R.~L.
  Pfefferbaum.
\newblock Community resilience as a metaphor, theory, set of capacities, and
  strategy for disaster readiness.
\newblock \emph{American Journal of Community Psychology}, 41\penalty0
  (1-2):\penalty0 127--150, 2008.

\bibitem[Ord(1975)]{ord75}
K.~Ord.
\newblock Estimation methods for models of spatial interaction.
\newblock \emph{Journal of the American Statistical Association}, 70\penalty0
  (349):\penalty0 120--126, 1975.

\bibitem[Pattison et~al.(2013)Pattison, Robins, Snijders, and Wang]{pattison13}
P.~E. Pattison, G.~L. Robins, T.~A.~B. Snijders, and P.~Wang.
\newblock Conditional estimation of exponential random graph models from
  snowball sampling designs.
\newblock \emph{Journal of Mathematical Psychology}, 57\penalty0 (6):\penalty0
  284--296, 2013.

\bibitem[Potterat et~al.(2004)Potterat, Woodhouse, Muth, Rothenburg, Darrow,
  Klovdahl, Muth, et~al.]{potterat04}
J.~Potterat, D.~E. Woodhouse, S.~Q. Muth, R.~B. Rothenburg, W.~W. Darrow, A.~S.
  Klovdahl, J.~B. Muth, et~al.
\newblock Network dynamism: history and lessons of the {Colorado Springs}
  study.
\newblock In \emph{Network epidemiology: A handbook for survey design and data
  collection}. Oxford University Press, 2004.

\bibitem[{R Core Team}(2013)]{R-manual}
{R Core Team}.
\newblock \emph{R: A Language and Environment for Statistical Computing}.
\newblock R Foundation for Statistical Computing, Vienna, Austria, 2013.
\newblock URL \url{http://www.R-project.org}.

\bibitem[Robins(2015)]{robins15}
G.~Robins.
\newblock \emph{Doing Social Network Research: Network-based Research Design
  for Social Scientists}.
\newblock Sage, London, 2015.

\bibitem[Robins et~al.(2001{\natexlab{a}})Robins, Elliott, and
  Pattison]{robins01}
G.~Robins, P.~Elliott, and P.~Pattison.
\newblock Network models for social selection processes.
\newblock \emph{Social Networks}, 23\penalty0 (1):\penalty0 1--30,
  2001{\natexlab{a}}.

\bibitem[Robins et~al.(2001{\natexlab{b}})Robins, Pattison, and
  Elliott]{robins01b}
G.~Robins, P.~Pattison, and P.~Elliott.
\newblock Network models for social influence processes.
\newblock \emph{Psychometrika}, 66\penalty0 (2):\penalty0 161--189,
  2001{\natexlab{b}}.

\bibitem[Robins et~al.(2004)Robins, Pattison, and Woolcock]{robins04}
G.~Robins, P.~Pattison, and J.~Woolcock.
\newblock Missing data in networks: exponential random graph ($p^\ast$) models
  for networks with non-respondents.
\newblock \emph{Social Networks}, 26\penalty0 (3):\penalty0 257--283, 2004.

\bibitem[Robins et~al.(2007)Robins, Snijders, Wang, Handcock, and
  Pattison]{robins07}
G.~Robins, T.~Snijders, P.~Wang, M.~Handcock, and P.~Pattison.
\newblock Recent developments in exponential random graph ($p^\ast$) models for
  social networks.
\newblock \emph{Social Networks}, 29\penalty0 (2):\penalty0 192--215, 2007.

\bibitem[Rothenberg et~al.(1995)Rothenberg, Woodhouse, Potterat, Muth, Darrow,
  and Klovdahl]{rothenberg95}
R.~B. Rothenberg, D.~E. Woodhouse, J.~J. Potterat, S.~Q. Muth, W.~W. Darrow,
  and A.~S. Klovdahl.
\newblock Social networks in disease transmission: the {Colorado Springs}
  study.
\newblock \emph{NIDA research monograph}, 151:\penalty0 3--19, 1995.

\bibitem[Sewell(2017)]{sewell17}
D.~K. Sewell.
\newblock Network autocorrelation models with egocentric data.
\newblock \emph{Social Networks}, 49:\penalty0 113--123, 2017.

\bibitem[Silk et~al.(2017)Silk, Croft, Delahay, Hodgson, Weber, Boots, and
  McDonald]{silk17}
M.~J. Silk, D.~P. Croft, R.~J. Delahay, D.~J. Hodgson, N.~Weber, M.~Boots, and
  R.~A. McDonald.
\newblock The application of statistical network models in disease research.
\newblock \emph{Methods in Ecology and Evolution}, 8\penalty0 (9):\penalty0
  1026--1041, 2017.

\bibitem[Smith and Moody(2013)]{smith13}
J.~A. Smith and J.~Moody.
\newblock Structural effects of network sampling coverage {I}: Nodes missing at
  random.
\newblock \emph{Social Networks}, 35\penalty0 (4):\penalty0 652--668, 2013.

\bibitem[Smith et~al.(2017)Smith, Moody, and Morgan]{smith17}
J.~A. Smith, J.~Moody, and J.~H. Morgan.
\newblock Network sampling coverage {II}: The effect of non-random missing data
  on network measurement.
\newblock \emph{Social Networks}, 48:\penalty0 78--99, 2017.

\bibitem[Snijders(2002)]{snijders02}
T.~A.~B. Snijders.
\newblock Markov chain {M}onte {C}arlo estimation of exponential random graph
  models.
\newblock \emph{Journal of Social Structure}, 3\penalty0 (2):\penalty0 1--40,
  2002.

\bibitem[Stivala et~al.(2016)Stivala, Koskinen, Rolls, Wang, and
  Robins]{stivala16}
A.~D. Stivala, J.~H. Koskinen, D.~A. Rolls, P.~Wang, and G.~L. Robins.
\newblock Snowball sampling for estimating exponential random graph models for
  large networks.
\newblock \emph{Social Networks}, 47:\penalty0 167--188, 2016.

\bibitem[Tange(2011)]{tange11}
O.~Tange.
\newblock Gnu parallel - the command-line power tool.
\newblock \emph{;login: The USENIX Magazine}, 36\penalty0 (1):\penalty0 42--47,
  Feb 2011.
\newblock URL \url{http://www.gnu.org/s/parallel}.

\bibitem[Thompson and Frank(2000)]{thompson00}
S.~K. Thompson and O.~Frank.
\newblock Model-based estimation with link-tracing sampling designs.
\newblock \emph{Survey Methodology}, 26\penalty0 (1):\penalty0 87--98, 2000.

\bibitem[Valente(2010)]{valente10}
T.~W. Valente.
\newblock \emph{Social networks and health: Models, methods, and applications}.
\newblock Oxford University Press, New York, 2010.

\bibitem[Wang et~al.(2009)Wang, Robins, and Pattison]{wang09}
P.~Wang, G.~Robins, and P.~Pattison.
\newblock \emph{PNet: program for the simulation and estimation of exponential
  random graph ($p^\ast$) models}.
\newblock Department of Psychology, The University of Melbourne, 2009.

\bibitem[Wang et~al.(2014)Wang, Neuman, and Newman]{wang14}
W.~Wang, E.~J. Neuman, and D.~A. Newman.
\newblock Statistical power of the social network autocorrelation model.
\newblock \emph{Social Networks}, 38:\penalty0 88--99, 2014.

\bibitem[Wickham(2009)]{wickham09}
H.~Wickham.
\newblock \emph{ggplot2: elegant graphics for data analysis}.
\newblock Springer, New York, 2009.
\newblock ISBN 978-0-387-98140-6.
\newblock URL \url{http://had.co.nz/ggplot2/book}.

\bibitem[Wilson(1927)]{wilson27}
E.~B. Wilson.
\newblock Probable inference, the law of succession, and statistical inference.
\newblock \emph{Journal of the American Statistical Association}, 22\penalty0
  (158):\penalty0 209--212, 1927.

\bibitem[Woodhouse et~al.(1994)Woodhouse, Rothenberg, Potterat, Darrow, Muth,
  Klovdahl, Zimmerman, Rogers, Maldonado, Muth, et~al.]{woodhouse94}
D.~E. Woodhouse, R.~B. Rothenberg, J.~J. Potterat, W.~W. Darrow, S.~Q. Muth,
  A.~S. Klovdahl, H.~P. Zimmerman, H.~L. Rogers, T.~S. Maldonado, J.~B. Muth,
  et~al.
\newblock Mapping a social network of heterosexuals at high risk for {HIV}
  infection.
\newblock \emph{AIDS}, 8\penalty0 (9):\penalty0 1331--1336, 1994.

\end{thebibliography}



\end{document}